\documentclass[
floatfix,
 reprint,
superscriptaddress,
nofootinbib,
 amsmath,amssymb,
 aps,
prc,
]{revtex4-2}

\usepackage{slashed}
\usepackage{physics}
\usepackage{amsmath}
\usepackage{graphicx}
\usepackage[inkscapelatex=false]{svg}
\usepackage{placeins} 
\usepackage{dcolumn}
\usepackage{bm}
\usepackage{hyperref}

\usepackage{lipsum}
\usepackage{color}
\usepackage{ulem}
\usepackage{comment}

\newcommand{\idp}{\int\!\frac{d^3p}{(2\pi)^3\!} }

\newcommand{\vac}{\eval_{\phi=(f_\pi,\Vec{0})}}

\newcommand{\sumint}{\displaystyle{\int} \kern-1em \textstyle{\sum}}

\renewcommand{\c}{\; ,}
\newcommand{\p}{\; .}

\newcommand{\JLU}{Institut f\"ur Theoretische Physik, Justus-Liebig-Universit\"at, Heinrich-Buff-Ring 16, 35392 Giessen, Germany}
\newcommand{\HFHF}{Helmholtz Forschungsakademie Hessen f\"ur FAIR (HFHF), GSI Helmholtzzentrum f\"ur Schwerionenforschung, Campus Giessen}

\newcommand{\TUD}{Technische Universität Darmstadt, 64289, Darmstadt, Germany}
\newcommand{\Urbana}{University of Illinois at Urbana-Champaign, Urbana, Il 61801, USA}

\begin{document}

\preprint{APS/123-QED}


\title{{Renormalization-Group Invariant Mean-field Analysis of the\\ Parity-Doublet Model for Nuclear and Neutron-Star Matter}}

\author{Mattia Recchi}\affiliation{\JLU}
\author{Lorenz von Smekal}\affiliation{\JLU}\affiliation{\HFHF}
\author{Jochen Wambach}\affiliation{\TUD}\affiliation{\Urbana}

\date{\today}

\begin{abstract}
The Parity-Doublet Model (PDM) is a chirally invariant effective theory for strong-interaction matter involving nucleons and their opposite-parity partners in a parity-doubling framework.
We introduce a multiplicatively renormalizable mean-field approach to include the baryonic vacuum contributions to the resulting grand-canonical potential in an explicitly renormalization-group invariant form. As an application, we evaluate the pertinent thermodynamics of two-flavor symmetric and asymmetric nuclear matter, focusing on the restoration of spontaneously broken chiral symmetry at baryon densities and temperatures relevant for the astrophysics of neutron stars. 
Special attention is paid to the effect of the baryonic vacuum fluctuations on the evolution of the chiral condensate with baryon density and temperature for specific choices of the chirally invariant baryon mass $m_0$  to demonstrate the importance of consistently including these vacuum fluctuations in the PDM.
\end{abstract}

\maketitle


\section{Introduction}

Understanding the structure of nuclear matter under extreme conditions is key to strong-interaction physics with applications in astrophysics and high-energy nuclear physics. As first-principles lattice-QCD (LQCD) calculations cannot be applied in the relevant regions of the phase diagram because of the fermion sign problem, modeling chiral-symmetry restoration through effective chiral theories is a central objective for the description of dense and hot baryonic matter realized in core-collapse supernovae, the interior of neutron stars (NS), in NS mergers, and in heavy-ion collisions at center-of-mass energies of a few GeV.

Chiral models, which have a long history \cite{Afonin:2007mj}, are based on the empirical observation that hadronic parity partners have different masses as a consequence of spontaneously broken chiral symmetry in the non-perturbative QCD vacuum. They reproduce the (nearly) exact degeneracy of the spectral properties of chiral partners of opposite parity
in the chirally restored phase as a general consequence of chiral invariance of the QCD Lagrangian for massless quarks. But there are also parity partners that are not chiral partners, i.e.~which are not directly related by chiral symmetry. This opens interesting possibilities for effective Lagrangians with chiral symmetry, especially for baryons.

For high baryon density, the parity-doublet model \cite{Detar:1988kn}, as a generalization of the linear sigma model \cite{Gell-Mann:1960mvl} and a chirally symmetric version of the Walecka model \cite{Serot:1984ey}, is particularly interesting. Although respecting chiral symmetry it involves massive baryonic parity partners with a common finite mass $m_0$ and only the mass splitting is generated by the chiral condensate. The mass $m_0$, the origin of which remains unspecified, serves as a parameter of the model. The Parity-Doublet Model (PDM) offers the intriguing possibility of a chirally restored phase with massive parity-doubled baryons \cite{Detar:1987kae,Detar:1987hib}. This phenomenon has been confirmed for baryon octet and decuplet screening masses in finite-temperature LQCD simulations \cite{Aarts:2017rrl}. In recent years, the PDM has gained attention for its application to nuclear saturation properties, the description of chiral  phase transitions in nuclear matter, and the behavior of the chiral condensate at finite density and in isolated NS \cite{Marczenko:2019trv,Kong:2023nue,Eser:2023oii}. Furthermore, extensions to flavor $SU(3)$ \cite{Steinheimer:2011ea,Minamikawa:2023eky,Fraga:2023wtd} have been made to study the role of strangeness in NS. Studies of baryon-number fluctuations in the PDM can be found in Refs.~\cite{Koch:2023oez, Marczenko:2023ohi,Marczenko:2024nge}, emphasizing the relevance for the experimental heavy-ion programs at FAIR/GSI energies \cite{Senger:2020pzs,HADES:2022gdr}.

The work described here builds on and extends previous two-flavor studies \cite{Eser:2023oii,Eser:2024xil} in a one-loop renormalized mean-field (MF) approximation. It presents a novel multiplicatively renormalizable treatment of the fermionic vacuum fluctuations, ensuring explicit renormalization group (RG) invariance of the grand-canonical potential. This procedure has proven quantitatively successful for pure-isospin QCD matter, where a direct comparison of the analogous RG-invariant MF calculation in the quark-meson model with LQCD results is possible \cite{Brandt:2025tkg}. 
In particular, the RG-invariant scale parameter $\Lambda $, here of the PDM, is generated by dimensional transmutation, as usual, to replace the renormalization scale and the relevant dimensionless interaction strength without introducing a new parameter.

The paper is organized as follows: In Sec.~\ref{sec:PDM} we recall the basic ideas of the PDM, especially emphasizing the role of the chirally invariant mass term $m_0$. After introducing the model for the two-flavor case, we discuss the resulting grand-canonical potential and the RG-invariant treatment of the divergent fermionic vacuum contribution (VC). We then present the strategy for fixing the model parameters through vacuum phenomenology and well-established properties of low-density symmetric and asymmetric nuclear matter. Sec.~\ref{sec:Results} is devoted to selected results on the phase structure of the chiral condensate as a function of the thermodynamic variables, temperature $T$, baryo-chemical potential $\mu_B$, and isospin-chemical potential $\mu_I$ for a range of chiral mass parameters $m_0$, contrasting the results with and without including the VC. Our particular focus is on the evolution of the chiral condensate in the vicinity of the liquid-gas (LG) transition of symmetric nuclear matter. We then turn to the astrophysical applications for $\beta$-equilibrated NS matter by discussing the composition, the Equation of State (EoS), and the resulting Mass-Radius (M-R)-relations for evolved NS together with their tidal deformabilities, as well as the thermal index which is relevant for proto-NS evolution and binary NS mergers. Sec.~\ref{sec:Conc} summarizes the pertinent findings of our paper and discusses future directions.

\section{The Parity-Doublet Model}\label{sec:PDM}

In this section we briefly review the main ideas of the two-flavor PDM, establishing the notation and discussing the resulting grand-canonical potential $\Omega(T,\mu_B,\mu_I)$. We then focus in detail on the RG-invariant treatment of the divergent fermionic VC to $\Omega$. The rest of the section is devoted to the choice of parameters and the fitting procedure.\footnote{We will use natural units $\hbar = c = k_B=1$ throughout.}

\subsection{The Model}
\label{sec:Model}

In its two-flavor version, the PDM is a generalization of the linear sigma model \cite{Gell-Mann:1960mvl} in which the sigma-meson and the pion fields are coupled to the nucleon $N(939)$ and its negative-parity partner, here assumed to be the $N^*(1535)$ resonance.
One starts with two sets of spin-$1/2$ baryons in two-flavor chiral $(\frac{1}{2},0) \oplus (0, \frac{1}{2}) $ representations, $N_1$ and $N_2$, with chirally invariant Yukawa couplings to the $O(4)$ vector $(\sigma,\vec \pi)$ of sigma-meson and pion fields.  Requiring the so-called ``mirror assignment'' \cite{Jido:1998av}, where the left-handed $N_2$ transforms as the right-handed $N_1$ and vice versa under $SU(2)_L\times SU(2)_R$, it is then possible to introduce a chirally invariant fermion mass $m_0 $, common to both parity-partner baryons, to model purely gluonic mass contributions from the QCD scale anomaly \cite{Yang:2018nqn,Burkert:2023wzr} which do not rely on chiral symmetry breaking.
The fermionic part $\mathcal{L}_F$ of the (Euclidean) Lagrangian is then given by 
\begin{equation}\label{eq:L_f}
\begin{split}
\mathcal{L}_F =&\bar{N}_1\big(\slashed\partial + g_1(\sigma + i\gamma_5 \vec\tau \cdot \vec\pi)\big)N_1 \\ 
&+ \bar{N}_2\big(\slashed\partial + g_2(\sigma - i\gamma_5 \vec\tau \cdot \vec\pi)
\big )N_2 \\
&+ m_0 \big(\bar{N}_1 \gamma_5 N_2 - \bar{N}_2 \gamma_5 N_1\big)  
\end{split}
\end{equation}
with the Yukawa couplings $g_1$ and $g_2$, where the opposite signs in the (pseudo-scalar) pion-baryon couplings reflect the mirror assignment of the chiral transformations of $N_1$ and $N_2$. With spontaneous chiral symmetry breaking, the $\sigma$ field receives a non-vanishing vacuum expectation value 
$\langle \sigma \rangle\equiv \sigma_0= f_\pi $, leading to additional Dirac mass contributions $m_1= g_1 f_\pi $  and $m_2= g_2 f_\pi $  to $N_1$ and $N_2$.
The physical positive $N_+=N(939)$ and negative $N_- = N^*(1535)$ parity baryon fields are then obtained from diagonalizing the mass matrix in the Nambu-Gorkov space of parity-partner baryons, so that chiral-symmetry breaking is responsible for the mass splitting of the two. 

For space-time independent, constant meson fields the diagonalization is straightforward, yielding 
\begin{equation}\label{eq:free_fermions}
\mathcal{L}_F =\bar{N}_+(\slashed\partial + m_+)N_+
+ \bar{N}_-(\slashed\partial + m_-)N_-  \c
\end{equation}
with the physical masses $m_\pm$ given by 
\begin{equation}\label{eq:masses}
m_\pm(\phi) =\frac{1}{2} \sqrt{(g_1+g_2)^2\phi^2 + 4m_0^2} \pm \frac{g_1-g_2}{2}\phi \, ,
\end{equation}
in terms of the constant radial $O(4)$-field variable $\phi^2 = \sigma^2 +\vec\pi^2  $. Of course, with the correct vacuum alignment, we may set $\phi \to \sigma $, but the form in Eq.~\eqref{eq:masses} explicitly reflects the fact that the fermion determinant is chirally invariant.

In the vacuum, with $\sigma_0 = f_\pi$, we use {$f_\pi = 93$~MeV for the pion decay constant,} $m_N  \equiv m_+(f_\pi)  = 939$~MeV for the nucleon and 
$ m_{N^*} \equiv m_-(f_\pi)  = 1510$~MeV for the $N^*(1535)$ as its parity partner \cite{ParticleDataGroup:2024cfk}. Since for any given value of $m_0 $: 
\begin{align}   
    g_2 - g_1 &= \frac{m_{N^*} - m_N}{f_\pi}  \approx 6.2 \, ,\label{eq:g1g2}\\
    g_1 + g_2 &= \frac{\sqrt{(m_N + m_{N^*})^2- 4 m_0^2}}{f_\pi} \, , \nonumber 
\end{align}
this then uniquely fixes $g_1$ and $g_2$.
As chiral symmetry gets restored in the medium, the sigma-field diminishes, either in a crossover transition or a first(second)-order phase transition. When chiral symmetry is fully restored ($\sigma=0$), according to Eq.~\eqref{eq:masses}, the parity-partner baryons become degenerate in mass with finite $m_0$.

In the mesonic sector, the MF approximation essentially amounts to adding an effective potential $U(\sigma,\vec\pi) $ to the Lagrangian, which is of the form 
\begin{equation}
    U(\sigma,\vec\pi) = V(\phi^2) - c\sigma \, ,
\end{equation}
with a polynomial in the $O(4)$-invariant $\phi^2$, conventionally starting with 
\begin{equation}
 V(\phi^2) =   - \frac{m^2}{2} \, \phi^2 + \frac{\lambda}{4}\, \phi^4 + \cdots \, . \label{eq:standardquaritcP}
\end{equation}
In the present work, we will use an $8^\mathrm{th}$-order polynomial for $V(\phi^2)$ of equivalent form but with slightly unusual coefficients, as we will detail below. 

The term $-c\sigma$ accounts for the explicit breaking of chiral symmetry due to finite $u$ and $d$ quark masses. In terms of the pion decay constant $f_\pi$ and the pion mass $m_\pi$, we have $c=f_\pi m_\pi^2$, here with $m_\pi = 135 ~\mathrm{MeV}$.
In the chirally broken vacuum only the $\sigma$ field is non-vanishing and equal to $f_\pi$. We will further assume that there is no pion condensation in nuclear and neutron matter and hence the classical pion field vanishes. Thus the $\sigma$-field is the only chiral order parameter throughout.
While the $\sigma$-field is treated classically, we will include fermionic vacuum fluctuations, as detailed in the next subsection, in a renormalized MF-approximation \cite{Skokov:2010sf,Brandes:2021pti,Eser:2023oii} which is equivalent to integrating the purely fermionic functional RG flow \cite{Weyrich:2015hha}. As we will see below, the inclusion of these fluctuations is crucial for the evolution of the $\sigma$-field and hence the restoration of chiral symmetry at high density.
 
\subsection{Grand-Canonical Potential}

In the path-integral representation of the grand-canonical partition function one includes chemical potentials $\mu_B $ for baryon number and $\mu_I$ for isospin imbalance (here chosen to be positive for neutron excess) by adding $- (\mu_B - \mu_I \tau^3 ) \gamma_0 $ to the diagonal blocks of the Nambu-Gorkov Dirac operator in Eq.~\eqref{eq:L_f}. We will also include the short-range repulsion between baryons in the form of four-Fermi interactions in the isoscalar and (neutral) isovector vector channels. {Assuming equal strengths of these for both, $N$ and $N^*$, chiral symmetry and parity then uniquely define the two allowed four-fermion interactions in the two channels \cite{Larionov:2021ycq}. These can be realized by including corresponding Hubbard fields $\omega_\mu $ and $\rho_\mu $ in the diagonal blocks of the Nambu-Gorkov Dirac operator, as well.} The repulsive nature implies pure imaginary couplings, replacing  
\begin{equation}
\slashed\partial\to\slashed\partial-i\gamma^\mu(g_\omega\omega_\mu - g_\rho\rho_\mu \tau^3) 
\end{equation}
in Eqs.~\eqref{eq:L_f} and/or \eqref{eq:free_fermions}, together with additional quadratic terms in the mesonic potential, defining
\begin{equation}
    U(\sigma,\vec\pi,\omega,\rho) = V(\phi^2) - c\sigma + \frac{m_\omega^2}{2}\, \omega^2   + \frac{m_\rho^2}{2}\, \rho^2 \, .   
\end{equation}
The strengths of the repulsive interactions are solely determined by the ratios $g_\omega^2/m_\omega^2 $ and $g_\rho^2/m_\rho^2 $, so that the assignment of   
the mass parameters is arbitrary. For comparison with conventional practice, we use the physical vector-meson masses, $m_\omega=783$~MeV and $m_\rho=776$~MeV, here. 
The MF approximation for the Hubbard fields then implies pure imaginary saddle points in their zero components, $\omega_0 = i\bar\omega$ and $\rho_0 = i\bar\rho $ with positive constants $\bar\omega, \, \bar\rho \ge 0$ {which amounts to a stationary-phase approximation with complex saddle points, from where the stable Lefschetz thimbles emerge parallel to the real axis.}
In the Nambu-Gorkov Dirac operator we can include these potentially non-vanishing expectation values {$\bar\omega$ and $\bar\rho$} by simple shifts in the corresponding chemical potentials, defining

\begin{equation}
    \tilde\mu_\tau = \mu_B - g_\omega \bar\omega - \tau  \big(\mu_I - g_\rho \bar\rho\big) \, , 
\end{equation}

\vspace*{.4cm}
\noindent
where we have introduced the coefficient $\tau = \pm 1$, with $\tau = +1 $ for the proton and its charged parity partner and $\tau = -1$ for the neutron and the neutral $N^*$, corresponding to the eigenvalues of the Pauli isospin matrices.

With these conventions, the grand-canonical MF potential of the two-flavor PDM can easily be derived to be of the form
\begin{widetext}  
\begin{equation}\label{eq:Omega}
	\Omega(T,\mu_B,\mu_I) =  -2T\sum_{\tau=\pm 1} \sum_\pm   \idp \Bigg( \ln \cosh \qty(\frac{E_\pm-\tilde\mu_\tau }{2T}) 
    +  \ln \cosh \qty(\frac{E_\pm+\tilde\mu_\tau}{2T}) \Bigg) + U(\sigma,\vec\pi,i\bar\omega,i\bar\rho) \c
\end{equation}
\end{widetext}
where $E_\pm \equiv \sqrt{p^2 + m_\pm^2(\phi)}$ are the meson-field dependent single-particle energies of $N$ and $N^*$, for which we can identify $\phi =\sigma $ for vanishing pion field.  

Focusing on the fermionic part of the potential we note that the momentum integration in Eq.~\eqref{eq:Omega} is ultraviolet divergent. 
We can separate finite parts from the logarithmically divergent ones, which are contained solely in the VC denoted by $\Omega_\mathrm{vac}$.
First we replace $\cosh(x)$ with $\cosh(|x|)$, in order to explicitly take out the $T \to 0^+$ limit of the fermionic potential:

\begin{equation}\label{eq:lncosh}
\begin{split}
\ln \cosh (\frac{\qty|E-\tilde\mu_\tau|}{2T}) =& \frac{\qty|E-\tilde\mu_\tau|}{2T} \\
& - \ln (1-f\qty(\frac{\qty|E-\tilde\mu_\tau|}{T})) \: ,
\end{split}
\end{equation}
with the Fermi-Dirac distribution $f(x) = {1}/{(e^x+1)}$.

Inserting Eq.~\eqref{eq:lncosh} into the grand-canonical potential, we obtain:
\begin{equation}
\Omega = \Omega_T + \Omega_n + \Omega_{\mathrm{vac}} + U \: ,
\end{equation}
with
\begin{equation}
\begin{split}
\Omega_T = 2T \sum_{\tau=\pm1}\sum_\pm&\idp \Bigg( \ln (1-f\qty(\frac{|E_\pm -\tilde\mu_\tau|}{T})) \\
&+ \ln (1-f\qty(\frac{|E_\pm +\tilde\mu_\tau|}{T})) \Bigg) \c \\
\end{split}
\end{equation}
\begin{align}
&\Omega_n =  2\sum_{\tau=\pm1}\sum_\pm \idp (E_\pm - |\tilde\mu_\tau|)\theta(|\tilde\mu_\tau| - E_\pm) \c \\
&\Omega_{\mathrm{vac}} = - 4\sum_\pm \idp \sqrt{p^2 + m_\pm^2(\sigma)} \p \label{Omega_vac}
\end{align}
Thus $\Omega$ separates into a thermal part $\Omega_T$ depending on temperature and chemical potential, which vanishes for $T=0$, a density-dependent zero-temperature contribution $\Omega_n$ from the Fermi sea, which vanishes for $\tilde\mu_\tau = 0$, and  $\Omega_\mathrm{vac}$, the VC containing the ultraviolet divergent parts, which needs to be regularized and renormalized as described in the next subsection.

Note that $\Omega$ is invariant under $\tilde\mu_\tau \to - \tilde\mu_\tau$. This implies that it is only a function of $|\tilde\mu_\tau|$. In the following, we consider $\tilde\mu_\tau>0$ without any loss of generality.

The Landau free-energy density as a function of $T$, $\mu_B$ and $\mu_I$ is obtained from the MF grand-canonical potential upon extremizing $\Omega$ with respect to the constant meson and Hubbard fields, i.e.~solving 

\begin{equation}
    \pdv{\Omega}{\sigma} = 0 \; , \quad
    \pdv{\Omega}{\bar\omega} = 0 \; , \quad
    \pdv{\Omega}{\bar\rho} = 0 \c
\end{equation}
at fixed $T$, $\mu_B$ and $\mu_I$.
The ultraviolet divergent contribution from $\Omega_\mathrm{vac}$, which is often omitted in MF calculations (in the so-called ``no-sea approximation''), is field dependent and thus relevant in the equation for $\sigma$.

\subsection{Vacuum Contribution}
We now analyze the VC denoted by $\Omega_\mathrm{vac}$.
Regularization and renormalization follow standard procedures 
\cite{Skokov:2010sf,Brandes:2021pti,Eser:2023oii}. Here we adopt an RG invariant formulation analogous to the one used in Ref.~\cite{Brandt:2025tkg} for the quark-meson model which provides an accurate description of the current LQCD data at vanishing $\mu_B$ but finite isospin density. Being a variant of the RG-MF formulation used previously, for convenience, we only briefly describe the main steps, adapted to the PDM where necessary. 
Readers less interested in these steps might directly  proceed with the resulting RG-invariant form of the grand-canonical potential $\Omega $ in Eq.~(\ref{eq:grandpot})

First, as frequently also used for Casimir forces, the ultraviolet divergent integral in Eq.~\eqref{Omega_vac} can be regularized analytically with 

\begin{equation}
\int p^2 dp  \, \big( p^2 + m^2)^{(1-\epsilon)/2} = \frac{m^{4-\epsilon } \sqrt{\pi}\,\Gamma\big(\frac{\epsilon -4}{2}\big)}{4 \Gamma\big(\frac{\epsilon -1}{2}\big) } \, . \label{eq:A1}
\end{equation}
For $\epsilon\to 0^+$ this yields:
\begin{align}
    \Omega_\mathrm{vac} =
 \frac{1}{\pi^2} \sum_\pm &\bigg( \frac{m_\pm^4}{4\epsilon} - \\
 & \frac{m_\pm^4}{16}\Big(1- 4\ln2 +4\ln \big(m_\pm/{\nu}\big) \Big) +\mathcal O(\epsilon) \bigg) \c \nonumber
\end{align}
where $m_\pm\equiv m_\pm(\sigma)$, and $\nu $ is the renormalization scale.
We see that the logarithmically divergent terms are proportional to 
\begin{align}
     \frac{1}{4} \big(m_+^4 + m_-^4 \big) &= \frac{m_0^4}{2} + (g_1^2+ g_2^2 - g_1g_2) m_0^2 \phi^2 \nonumber\\
    & \hskip .6cm + \frac{1}{4} (g_1^4+g_2^4 )\phi^4 \p \label{eq:quarticP}
\end{align}
This implies that we can adopt the standard MF renormalization of the celebrated Gell-Mann-Levy model by 
introducing counter-terms of precisely this same structure, differing from the quartic meson self-interactions $\propto \phi^4 $ only by the term $\propto m_0^2 \phi^2 $ and an irrelevant (field-independent) constant.
We therefore \emph{define} the relevant terms of the tree-level mesonic potential for a multiplicatively renormalizable MF theory of the PDM by
\begin{align}
V(\phi^2) \equiv &  - \frac{1}{2} \frac{m^2}{g_1^2+ g_2^2}  \qty(m_+^2(\phi) + m_-^2(\phi) )   \nonumber \\ 
 &\hskip .2cm + \frac{\lambda}{4(g_1^4+g_2^4)} \qty(m_+^4(\phi) +  m_-^4(\phi))  \p \label{eq:mesonic-potential}
\end{align}
Since 
\begin{align}
    \frac{1}{2} \big(m_+^2 + m_-^2\big)  &=  m_0^2 + \frac{1}{2} (g_1^2+g_2^2 ) \phi^2 \c 
\end{align}
the first term in Eq.~\eqref{eq:mesonic-potential}  yields the same quadratic term as in Eq.~\eqref{eq:standardquaritcP}. Therefore Eq.~\eqref{eq:mesonic-potential} only differs from the conventional form in Eq.~\eqref{eq:standardquaritcP} by the additional term $\propto m_0^2 \phi^2 $ from Eq.~\eqref{eq:quarticP}.

Beyond the quartic polynomial in Eq.~\eqref{eq:mesonic-potential}, which contains the relevant and marginal couplings (allowed by symmetry), we can add irrelevant couplings $c_{2n}$ in higher order interaction terms of the analogous form
\begin{equation}
V_{2n}(\phi^2) \equiv    \frac{c_{2n}}{2n\,  m_0^{2n-4}} \left( m_+^{2n}(\phi) +  m_-^{2n}(\phi) \right) \c  
\end{equation}
for arbitrary powers $n\ge 3 $ in the mesonic field invariant $\phi^2$, noting that these all produce (RG-invariant) polynomials of order $n$ in $\phi^2$
whose leading behavior at large fields is given by
\begin{equation}
V_{2n}(\phi^2) \, = \,    \frac{c_{2n}}{2n \, m_0^{2n-4}}  \,   (g_1^{2n}+g_2^{2n} )   \phi^{2n} \, + \mathcal O\big( \phi^{2n-2}\big) \, .
\end{equation}
In order to optimize the quantitative description of the thermodynamic properties of nuclear and neutron matter, as described in the next section, here we include the dimensionless higher-order couplings $c_6$ and $c_8$, corresponding to $n=3$ and $4$ with
\begin{align}
V_6(\phi^2) = &   \frac{c_6}{6m_0^2} \left( m_+^6(\phi) +  m_-^6(\phi) \right) \c  \\
V_8(\phi^2) = &   \frac{c_8}{8m_0^4} \left( m_+^8(\phi) +  m_-^8(\phi) \right)  \, .
\end{align}
{The $n=3$ term} allows {for a first order LG transition that would otherwise not be possible after the inclusion of the VC, while the $n=4$ term improves the description of the LG critical point}

{We note that this expansion of the mesonic potential represents a truncation of an (infinite) series in even powers of the masses, and hence the meson fields (as dictated by chiral symmetry), effectively including quantum corrections and many-body correlations. For simplicity, we decided to keep this truncation at a minimal number of terms that satisfy the physical constraints. We acknowledge that this can lead to a globally unbounded mesonic potential. Stability can always formally be restored by including higher-order irrelevant couplings $c_{2n}  $ for $n>4 $ in the truncation that will have numerically negligible effects on the results as can be judged from the clear hierarchy in the magnitudes of the dimensionless coefficients $c_6$ and $c_8$ as provided in Section III below. 
However, within the MF framework employed here, mesonic fluctuations are entirely neglected. Consequently, the dynamical mechanisms to trigger false vacuum decay are absent and global stability is not required.} 

Expressing the mesonic potential in this way, we can focus on the quartic term in \eqref{eq:mesonic-potential} for the 
renormalization of the logarithmic divergences from the vacuum contribution. In particular, we define the effective coupling  
\begin{equation} \label{eq:defu}
u \equiv     \frac{g_1^4 + g_2^4}{\lambda} \, ,
\end{equation}
which controls the MF approximation in the PDM, in a way analogous to $g^4/\lambda $ in the procedure of Ref.~\cite{Brandt:2025tkg} for the quark-meson model.
The renormalization is then achieved by noting that the pre-factors of all terms proportional to $m_\pm^4/4 $ in $\Omega$ can be combined to yield,
\begin{align}
  \frac{1}{u} + \frac{1}{\pi^2} \bigg( \frac{1}{\epsilon} + \ln 2 -  \frac{1}{4} -  \ln \big(m_\pm/{\nu}\big) \Big) \bigg) &\equiv \\
  & \hskip -2cm  
  \frac{1}{u_R(\nu) } - \frac{1}{\pi^2} \ln \big(m_\pm/{\nu}\big) \, ,\nonumber
\end{align}
where $ u_R(\nu) = Z_u^{-1}  u $ is the renormalized effective coupling at the renormalization scale $\nu$, and
\begin{equation}
    Z_u = 1 + \frac{ u}{\pi^2 } \Big( \frac{1}{\epsilon}  + \ln 2 - \frac{1}{4} \Big)\, .
\end{equation}
Note that, in absence of pion condensation for sufficiently small $\mu_I$, which is the case considered here, it is in principle sufficient to renormalize the quartic meson-coupling, $\lambda_R = Z_\lambda^{-1}\lambda $ with $Z_\lambda^{-1} = Z_u $. Due to the need of meson-field renormalization in the seagull terms, in the presence of pion condensation, this is not sufficient in general, however. As in Ref.~\cite{Brandt:2025tkg}, the complete multiplicative MF renormalization, is then defined by $\phi_R = Z_\phi^{-1/2} \phi $ for the meson fields, $m_R^2 = Z_m^{-1} m^2$ for their mass parameter, and with equal renormalization factors 
 $g_{1,R}^2 = Z_g^{-1} g_1^2$,  $g_{2,R}^2 = Z_g^{-1} g_2^2$, for both Yukawa couplings. One can then conveniently choose $Z_u = Z_g^2/Z_\lambda $ and $Z_g $ as the two independent renormalization factors that are needed, with the others determined by $Z_\phi Z_g = Z_\phi Z_m = 1$ \cite{Brandt:2025tkg}. Matching powers of meson fields and Yukawa couplings, such as $(g_1^2 + g_2^2) \phi^2 $, are then RG-invariant, and so is the ratio 

 \begin{equation} \label{eq:defmt}
 \tilde m^2 \equiv \frac{m^2}{g_1^2+g_2^2}
 \end{equation}
 which determines the mass term in the mesonic potential in Eq.~\eqref{eq:mesonic-potential}. With the RG-invariant mass parameter $m_0 $ of the PDM, the physical baryon masses $m_\pm(\sigma) $ are then automatically RG invariant as well, as they must.

The total effective potential for the sigma and pion fields, including the VC, can then be collected into
\begin{align}
    V_\mathrm{eff}(\phi) &= \sum_\pm \bigg\{ - \tilde m^2 \, \frac{m_\pm^2(\phi)}{2}  \label{eq:Veff} \\
    &\hskip 1cm + \bigg( \frac{1}{u_R(\nu)} -\frac{1}{\pi^2} \ln\frac{m_\pm(\phi)}{\nu} \bigg)\,  \frac{m_\pm^4(\phi)}{4} \nonumber\\
    &\hskip 3cm + c_6  \, \frac{m_\pm^6(\phi)}{6 m_0^2} +  c_8 \, \frac{m_\pm^8(\phi)}{8 m_0^4} \bigg\}\, 
    \nonumber
\end{align}
which is clearly RG invariant with $\beta(u) = u^2/\pi^2 $. Since it is renormalized and finite, we drop the index $R$ again, denoting by $u \equiv u_R(\nu)$ the scale-dependent renormalized effective coupling from now on. Introducing the scale parameter $\Lambda $ of the RG-invariant MF model via
   $ \Lambda \equiv \nu \, e^{\pi^2/u} $,
we can explicitly eliminate the remaining RG-scale $\nu $ dependence in Eq.~\eqref{eq:Veff} by the replacement
\begin{equation}
    \frac{1}{u} - \frac{1}{\pi^2}\ln\frac{m_\pm(\phi)}{\nu} = - \frac{1}{\pi^2}\ln\frac{m_\pm(\phi)}{\Lambda} \, .
\end{equation}
Alternatively, we may introduce the effective coupling $u_0 = u(\nu_0) $, using the chirally-invariant PDM mass $m_0$ as the reference scale, $\nu_0 = m_0$, to define our renormalization scheme. This then leads to an explicitly RG-scale independent form of the grand-canonical potential $\Omega$ as follows:
\begin{widetext}
    \begin{align}
\Omega(T,\mu_B,\mu_I) =& 
\sum_\pm \bigg\{ - \tilde m^2 \, \frac{m_\pm^2(\phi)}{2} + \frac{1}{\pi^2} \Big( \ln\frac{\Lambda}{m_0} - \ln\frac{m_\pm(\phi)}{m_0} \Big)\,  \frac{m_\pm^4(\phi)}{4}  + c_6  \, \frac{m_\pm^6(\phi)}{6 m_0^2} +  c_8 \, \frac{m_\pm^8(\phi)}{8 m_0^4} \bigg\}
\nonumber \\
& -c\sigma - \frac{m_\omega^2}{2}\bar\omega^2 - \frac{m_\rho^2}{2}\bar\rho^2 + \frac{2}{\pi^2}\sum_{\tau=\pm1}\sum_\pm \int p^2 dp \, (E_\pm - |\tilde\mu_\tau|)\theta(|\tilde\mu_\tau| - E_\pm) 
\label{eq:grandpot} \\
&+ \frac{2T}{\pi^2} \sum_{\tau=\pm1}  \sum_\pm \int p^2 dp \ \ln (1-f\qty(\frac{|E_\pm-\tilde\mu_\tau|}{T})) +\ln (1-f\qty(\frac{|E_\pm+\tilde\mu_\tau|}{T})) \p \nonumber
    \end{align}
\end{widetext}

{We emphasize that the results concerning the $\beta$-function and the subsequent RG invariance hold within the framework of purely fermionic fluctuations. Extending the analysis beyond mean-field by incorporating the full set of quantum fluctuations, including mesonic ones, would introduce additional contributions. Such effects are assumed to be effectively encoded in higher-order terms in the mesonic potential.}

Since, in this scheme, we have $m_0 = \Lambda \, e^{-\pi^2/u_0}$, we need $\Lambda \gg m_0 $
for any reasonably small and positive $u_0 $ (choosing $\Lambda = m_+(f_\pi) = m_N  $ would correspond to the scheme used in Ref.~\cite{Eser:2023oii}). Here, we will use the parameters $\tilde m^2 = {m^2}/{(g_1^2+ g_2^2)}$ and $\tilde\lambda_0 = 1/u_0 = \ln(\Lambda/m_0)/\pi^2$ to fix $\sigma_0 = f_\pi $ (at the reference scale) and the $\sigma$-meson (curvature) mass parameter $m_\sigma $, see below.   

\subsection{Parameter Fixing and Best Fit}
We now focus on the choice of the parameters for the PDM. These are fixed both from hadron phenomenology and the saturation properties of nuclear matter. 



The distinctive parameter of the PDM is the chirally invariant mass $m_0$ which enters the physical masses of the parity partners from Eq.~\eqref{eq:masses}, with
\begin{equation}
m_+ \qty(f_\pi) = m_{N} \, , \;\;  
m_- \qty(f_\pi) =  m_{N^*} \, ,
\end{equation} 
where for any given $m_0$ the two Yukawa couplings $g_1$ and $g_2$ are chosen such that the vacuum masses of the nucleon and the $N^*(1535)$ are given by $m_N= 939$~MeV and $m_{N^*}=1510$~MeV as described in Sec.~\ref{sec:Model}. The value of the chirally invariant baryon mass $m_0$ is rather poorly constrained, however. 
On physical grounds, we expect it to vary roughly between
\begin{equation}
500\;\mbox{MeV} \leq m_0\leq 860\;\mbox{MeV}\p
\end{equation}
The upper value results from the QCD trace anomaly in the chiral limit of vanishing light-quark masses \cite{Yang:2018nqn,Burkert:2023wzr}, while $m_0$ between $500$~MeV and $550$~MeV has been estimated from QCD sum rules \cite{Kim:2021xyp}. 


To bracket these possibilities we chose a range of $m_0$ between 500~MeV and 800~MeV in the following. For any fixed value of $m_0$ the parameters in the baryonic Lagrangian \eqref{eq:L_f} are thus fixed from Eqs.~\eqref{eq:g1g2}. 

The leading mesonic couplings are $ m^2$ and $\lambda$, or equivalently, $\tilde{m}^2$ and $u$ as defined in Eqs.~\eqref{eq:defmt} and \eqref{eq:defu}. While $\tilde m^2$ is RG invariant, the latter is RG-scale $\nu $ dependent.  Instead of $u \equiv u(\nu)$, we therefore use the RG invariant scale parameter $\Lambda = \nu \, e^{\pi^2/u} = m_0\, e^{\pi^2/u_0 }$ in our scheme, as introduced above.
These two  are fixed in the vacuum, requiring the expectation value of the $\sigma$-field and its Euclidean (curvature) mass parameter at the reference scale
to be given by $f_\pi$ and $m_\sigma^2$,\footnote{In general, these relations involve finite renormalization factors from meson field renormalization.}
\begin{equation}
\begin{split}
\dv{\Omega}{\sigma}\vac &= 0 \c \\
\dv[2]{\Omega}{\sigma}\vac &= m_\sigma^2 \p \\
\end{split}
\end{equation}
Since the mass parameter of the $\sigma$-meson as a broad two-pion resonance, is not very well constrained phenomenologically, we allow it to vary between 400~MeV and 550~MeV. It is otherwise not included in the fit procedure described next, but listed as a result, separated by a double line in the last column of Tab.~\ref{tab:fit_prms}.

In addition to $\Lambda $ and $\tilde m^2$, the remaining parameters that need to be fixed are given by the dimensionless mesonic $c_6$ and $c_8$ in \eqref{eq:Veff} and the strengths $g_\omega $ and $g_\rho$ of isoscalar and isovector vector repulsion between the baryons.   
To achieve this we use a best-fit procedure by minimizing the residual function 
\begin{equation}
\label{eq:res}
    \mathrm{Res}(\vec{x}) = \sum_i\frac{(x^\mathrm{exp}_i-x_i)^2}{\sigma_i^2} \, ,
\end{equation}
where $\vec{x}^\mathrm{\, exp}$ is the vector of phenomenological quantities with associated uncertainties $\vec{\sigma}$. 
The minimization is performed through a global-local hybrid approach: we use the DIRECT algorithm to search for a region containing the global minimum and then refine the solution using a standard Levenberg–Marquardt algorithm.

The quantities that enter in the fit are the nuclear saturation density, $n_0$, the binding energy per nucleon of symmetric nuclear matter, $E_B$, its compressibility at the saturation point, $K_\infty$, the symmetry energy, $E_\mathrm{sym}$, and the value of the in-medium chiral condensate $\sigma(n_0)$, deduced from other sources.

Following the recent analysis in \cite{Drischler:2024ebw}, $n_0$  and $E_B$ are taken as
\begin{equation}
\begin{cases}
n_0 =0.157 \pm 0.010 \text{ fm}^{-3} \, , \\

\vspace{-2 mm} \\

E_B=15.97 \pm 0.40 \text{ MeV} \p
\end{cases}
\end{equation}

The compressibility is defined as
\begin{equation}
K_\infty = 9 n_0^2 \pdv[2]{E}{n}\bigg|_{n_0} \c
\end{equation}
where $E$ is the energy per particle of the system. The value of $K_\infty$ is around $240 \pm 20$~MeV \cite{Stone:2014wza}.

To fix the value of $g_\rho$ we need a quantity related to the isospin asymmetry parameter $\delta=(n_n-n_p)/(n_p+n_n)$ where $n_n$ and $n_p$ are the neutron and proton densities, respectively. The symmetry energy is then given by
\begin{equation}
    E_\mathrm{sym} = \frac{1}{2} \pdv[2]{E}{\delta}\bigg|_{n_0, \delta=0} \p
\end{equation}
We take $E_\mathrm{sym}$ as $31 \pm 2$~MeV \cite{Baldo:2016jhp}. 

Since one of the main focuses of our study is the dynamics of the in-medium chiral condensate, we have decided to include in the parameter-fixing procedure also a quantity related to that, namely the value of the chiral condensate at saturation density. A natural way to introduce such a constraint is through the nucleon sigma term, $\sigma_N$, which measures the explicit chiral symmetry breaking contribution to the nucleon mass. In particular, $\sigma_N$ is defined as the nucleon matrix element of the light quark mass term,
\begin{equation}
\sigma_N = m_q \langle N | \bar{u}u + \bar{d}d | N \rangle = m_q\pdv{m_N}{m_q}\c
\end{equation}
and provides a measure of the sensitivity of the nucleon mass to changes in the quark masses.
It also governs the leading linear density dependence of the in-medium condensate. In fact, using the Feynman-Hellman theorem and the Gell-Mann-Oakes-Renner relation, one can relate the reduction of the condensate at finite density to the value of $\sigma_N$, leading to the expression \cite{Cohen:1991nk}:

\begin{equation}\label{eq: in_medium_sigma}
\sigma (n_0) = f_\pi \qty( 1-\frac{\sigma_N n_0}{f_\pi^2 m_\pi^2}) \c
\end{equation}
where $n_0$ denotes the saturation density and $f_\pi$, $m_\pi$ are the pion decay constant and mass, respectively.

The physical value of $\sigma_N$ is still somewhat controversial, with lattice simulations providing values roughly around $44 \pm 4$~MeV \cite{Agadjanov:2023efe,Alexandrou:2024ozj}, while the analysis of $\pi N$-scattering yields a value of about $56 \pm 4$~MeV, without isospin-breaking effects \cite{Hoferichter:2015dsa,Hoferichter:2023ptl} (also see the compilation in 
Ref.~\cite{Owa:2023tbk} which reports a value between the two).

Moreover, it is well established that these vacuum values lead to an unrealistically strong in-medium reduction of the chiral condensate once higher-order corrections 
are considered. The saturation density might not be small enough for the linear approximation in \eqref{eq: in_medium_sigma} to be valid.
To include these higher-order effects,
we follow Ref.~\cite{Kaiser:2007nv} which can be used to provide a suitable \textit{effective in-medium sigma term} at saturation density of $\tilde\sigma_N(n_0) = 35 \pm 8$~MeV 
to incorporate the non-linear corrections to Eq.~\eqref{eq: in_medium_sigma}, where the central value is based on $\sigma_N = 45$~MeV, and hence consistent within current errors with the recent lattice estimates of the standard nucleon-sigma term  at vanishing density.

\begin{table*}
    \centering
    \textbf{No Vacuum Contribution}\\
    \vspace{1 mm}
    {
    \begin{tabular}{|c||c|c|c|c|c|c||c|}
    \hline
$m_0$[MeV] 
& $\Lambda$ [MeV] & $\tilde{m}^2/\Lambda^2$ & $c_6$ & $c_8$ & $g_\omega$ &$g_\rho$& $m_\sigma$[MeV] \\
        \hline
        500 & 515.49 & $9.09207 \times 10^{-3}$ & $-3.05446 \times 10^{-4}$ & $1.04577 \times 10^{-5}$ & 8.27788 & 8.13045 & 528.872 \\
        600 & 620.40 & $6.85587 \times 10^{-3}$ & $-4.84265 \times 10^{-4}$ & $2.38632 \times 10^{-5}$ & 7.44413 & 8.24211 & 486.085 \\
        700 & 731.27 & $6.17614 \times 10^{-3}$ & $-9.10951 \times 10^{-4}$ & $6.41217 \times 10^{-5}$ & 6.34995 & 8.36055 & 440.269 \\
        800 & 880.89 & $8.47622 \times 10^{-3}$ & $-2.93592 \times 10^{-3}$ & $2.95590 \times 10^{-4}$ & 4.83758 & 8.48204 & 418.376 \\
        \hline
    \end{tabular}}\\
        \vspace{1 mm}
        \textbf{Vacuum Contribution}\\
        \vspace{1 mm}
    {
    \begin{tabular}{|c||c|c|c|c|c|c||c|}
    \hline
        $m_0$[MeV] &
        $\Lambda$ [MeV] & 
        $\tilde{m}^2/\Lambda^2$ &
        $c_6$ & $c_8$ & $g_\omega$ &$g_\rho$& $m_\sigma$[MeV] \\
        \hline
    500 & 1357.40 & $1.65563 \times 10^{-2}$ & $6.82367 \times 10^{-3}$ & $-1.51952 \times 10^{-4}$ & 7.02557 & 8.07853 & 475.006 \\
    600 & 1389.28 & $1.65804 \times 10^{-2}$ & $9.41046 \times 10^{-3}$ & $-2.91054 \times 10^{-4}$ & 6.50977 & 8.18026 & 451.000 \\
    700 & 1422.34 & $1.66018 \times 10^{-2}$ & $1.22525 \times 10^{-2}$ & $-4.96319 \times 10^{-4}$ & 5.91156 & 8.38832 & 400.000 \\
    800 & 1514.21 & $ 1.64299 \times 10^{-2}$ & $1.39845 \times 10^{-2}$ & $-6.33925 \times 10^{-4}$ & 4.75868 & 8.21138 & 400.002 \\
        \hline
    \end{tabular}
}
    \caption{Model parameters obtained from fitting the phenomenological properties of nuclear matter in Tab.~\ref{tab:fit_quantities} without (top) and with the VC (bottom) for different values of fixed $m_0$, from left: the scale parameter $\Lambda $ (without the VC simply defined as $\Lambda = m_0 \, \exp\{\pi^2 \lambda/(g_1^4+g_2^4)\}$), the mass parameter $\tilde m^2 = m^2/(g_1^2+g_2^2)$ in the mesonic potential \eqref{eq:mesonic-potential}, and the dimensionless parameters $c_6$, $c_8$, $g_\omega$, $g_\rho $ for the strengths of higher-order couplings and vector repulsion ($m_\sigma$ is allowed to vary between 400~MeV and 550~MeV, but otherwise not constrained and quoted for reference here, see text).     }
    \label{tab:fit_prms}
\end{table*}

\begin{table*}
    \centering
    \textbf{No Vacuum Contribution}\\
    \vspace{1 mm}
    \begin{tabular}{|c||c|c|c|c|c||c|c|}
    \hline
        $m_0$ [MeV] & $E_B$ [MeV]  & $n_0$ [fm$^{-3}$] & $E_\mathrm{sym}$ [MeV] & $K_{\infty}$ [MeV] & $\sigma(n_0)$ [MeV] & $L$ [MeV] & $T_c$ [MeV]\\
            \hline
            500 & 16.0 & 0.157 & 31.0  & 240 & 67.19  & 84 & 17.2\\
            600 & 16.0 & 0.157 & 31.0  & 240 & 67.19  & 83 & 17.3\\
            700 & 16.0 & 0.157 & 31.0  & 240 & 67.19  & 82 & 18.0\\
            800 & 16.0 & 0.157 & 31.0  & 240 & 67.19  & 80 & 21.7\\
            \hline
    \end{tabular}\\
    \vspace{1 mm}
    \textbf{Vacuum Contribution}\\
    \vspace{1 mm}
    \begin{tabular}{|c||c|c|c|c|c||c|c|}
        \hline
         $m_0$ [MeV] & $E_B$ [MeV]  & $n_0$ [fm$^{-3}$] & $E_\mathrm{sym} $ [MeV] & $K_{\infty}$ [MeV] & $\sigma(n_0)$ [MeV] &$L$ [MeV] & $T_c$ [MeV]\\
        \hline
        500 & 15.9 & 0.162 & 30.99 & 244.0 & 72.12 & 82.61 & 14.2\\
        600 & 15.8 & 0.161 & 31.01 & 244.0 & 71.21 & 82.08 & 15.4\\
        700 & 16.0 & 0.157 & 30.95 & 240.0 & 69.48 & 81.49 & 16.4\\
        800 & 15.8 & 0.164 & 31.01 & 259.0 & 66.73 & 80.47 & 21.6\\
        \hline
    \end{tabular}
    \caption{
    Phenomenological properties of nuclear matter near saturation as obtained from the best-fit parameters in Tab.~\ref{tab:fit_prms} without (top) and with the VC (bottom) for matching values of $m_0$, from left: the binding energy $E_B$ per nucleon of symmetric nuclear matter, its saturation density $n_0$,
    the symmetry energy $E_\mathrm{sym}$, the compressibility $K_\infty$ of infinite nuclear matter, and the in-medium chiral condensate $\sigma(n_0)$ at saturation density. The  slope parameter $L$ and the LG critical temperature $T_c$ are not used in the best-fit procedure, but are quoted here as resulting model predictions.}
    \label{tab:fit_quantities}
\end{table*}

Plugging $\tilde \sigma_N (n_0)$ into Eq.~\eqref{eq: in_medium_sigma} (in the place of $ \sigma_N$) we obtain:
\begin{equation}
\sigma(n_0) = 68 \pm 15 \,\text{MeV}.
\end{equation}

The best-fit parameters are shown in Tab.~\ref{tab:fit_prms} together with the resulting $\sigma$-meson mass parameter $m_\sigma $. The parameter sets are labeled by the chiral baryon mass parameter $m_0$ used as an input. 
The second column shows the RG-invariant scale parameter $\Lambda $ implicitly defined by the inverse of the quartic meson coupling $\lambda$ at the reference scale $\nu = m_0$ in our scheme as explained above. In the no-sea approximation (labeled ``No Vacuum Contribution'') the effective coupling $u = (g_1^4+g_2^4)/\lambda $ is of course not scale dependent, and we simply define  $\Lambda \equiv m_0\, e^{\pi^2/u}$ as a proxy for $\lambda $ for a better comparison in this case. The third column lists the meson mass parameter $\tilde m^2 = m^2/(g_1^2+g_2^2) $ in units of $\Lambda $. These two are essentially determined from $f_\pi $ (not shown in the table) and $m_\sigma $.  The remaining dimensionless parameters $c_6$, $c_8$, $g_\omega$ and $g_\rho $ of Tab.~\ref{tab:fit_prms} then lead to the physical quantities listed in  Tab.~\ref{tab:fit_quantities}.\footnote{{The negative values of $c_8$ are irrelevant on the MF level as discussed in Section II above. The mesonic potential always has a stable minimum at $\sigma = f_\pi$ without mesonic fluctuations.}}

\begin{figure*}
    \includegraphics[width=0.45\linewidth]{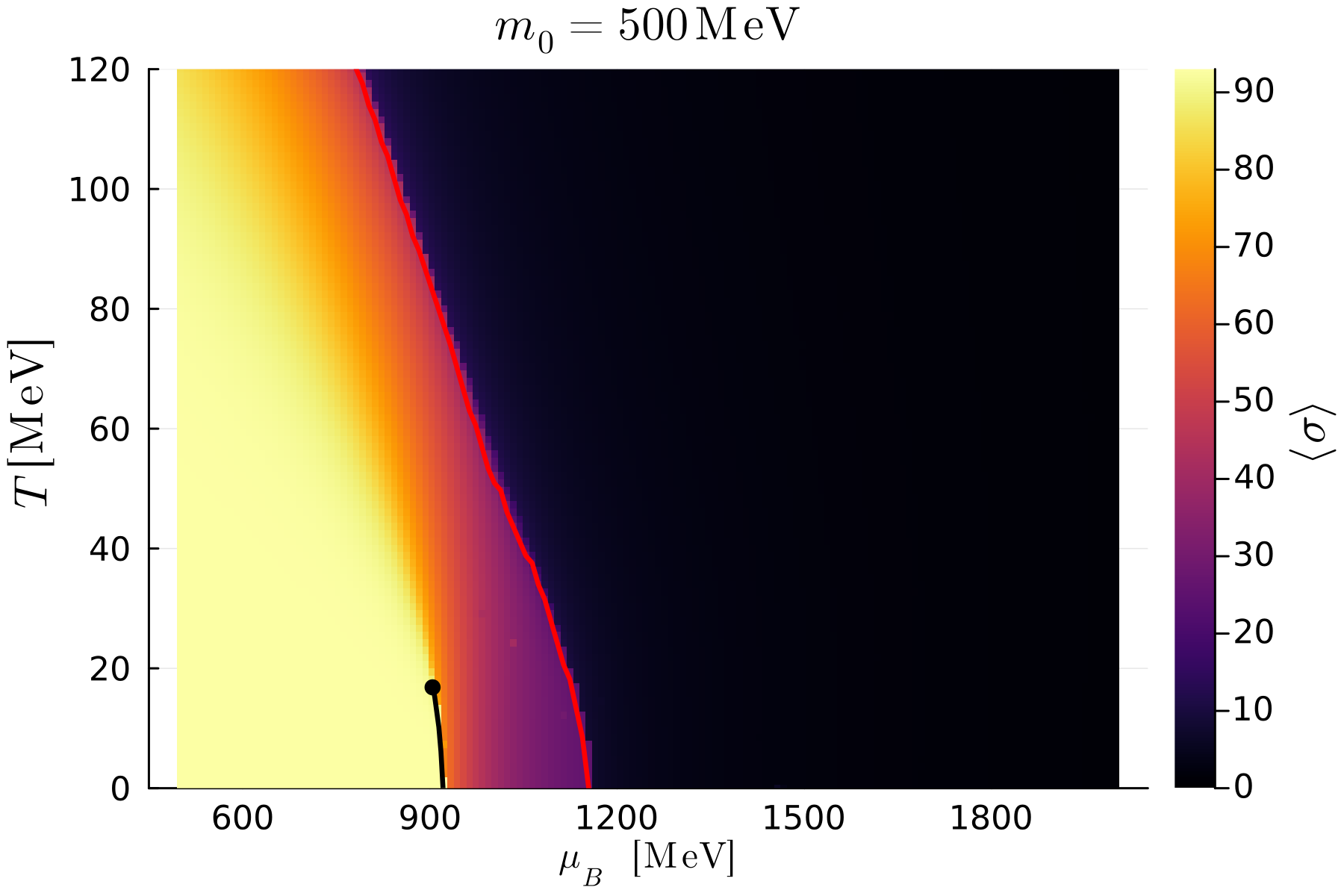}
    \includegraphics[width=0.45\linewidth]{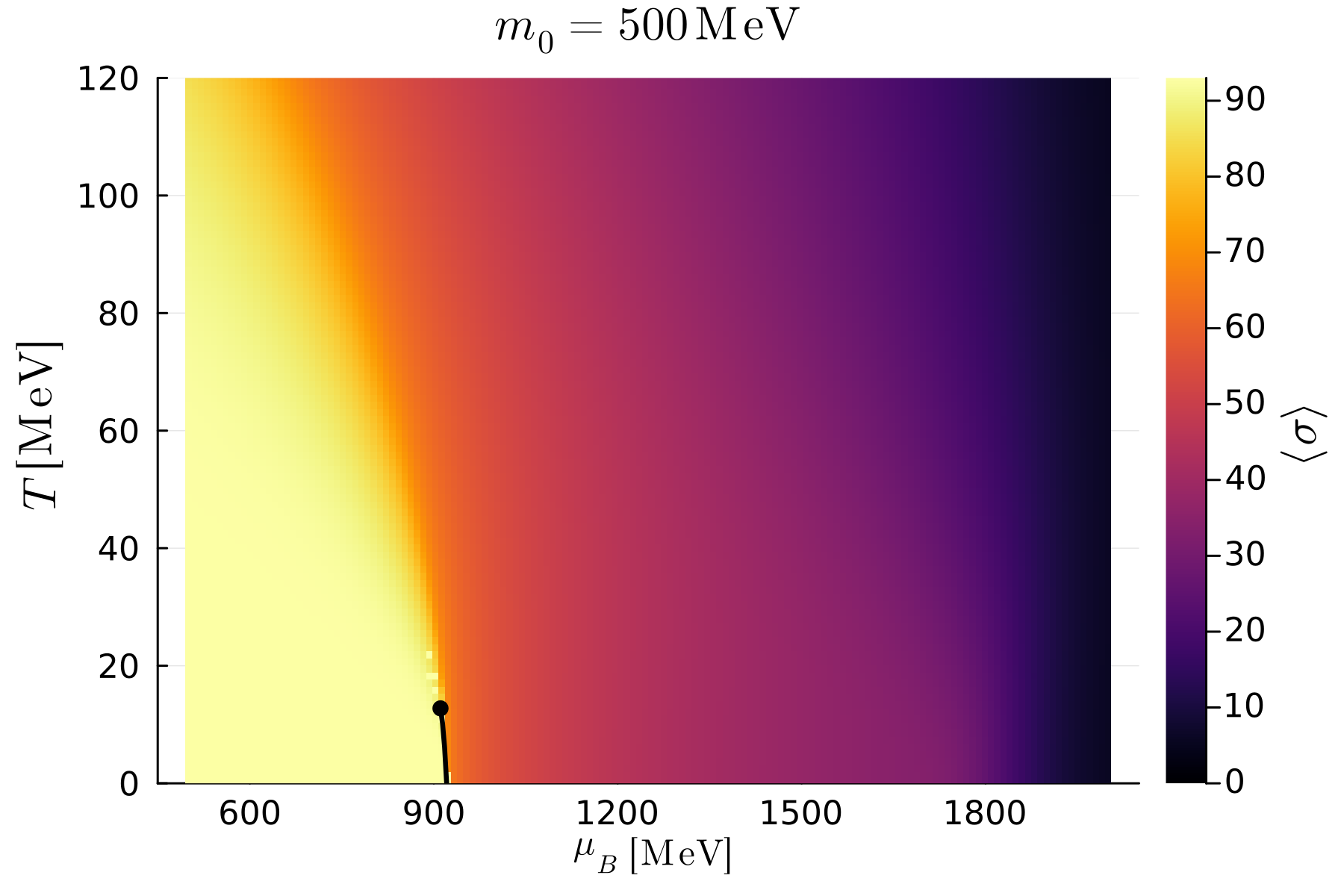}

    \includegraphics[width=0.45\linewidth]{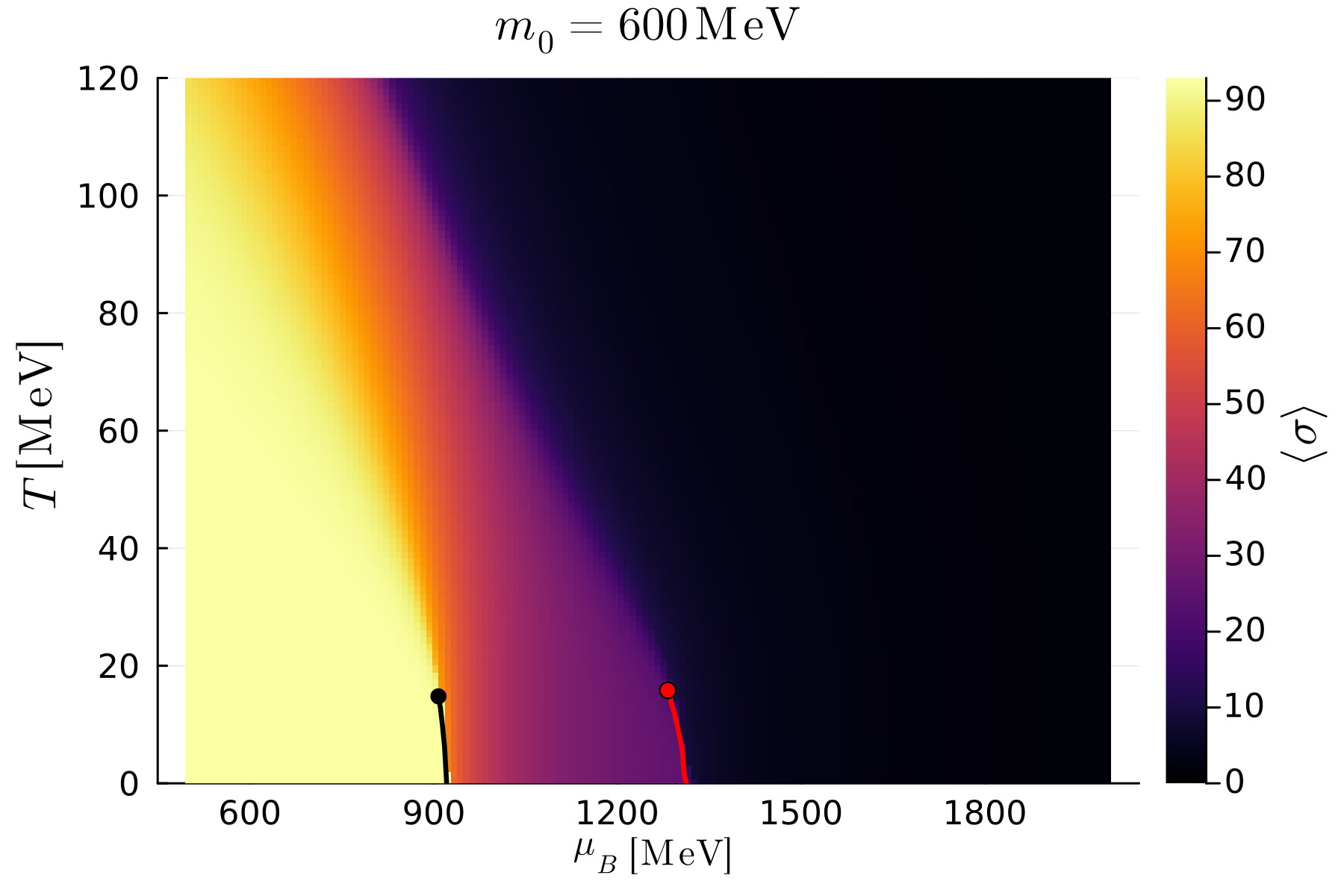}
    \includegraphics[width=0.45\linewidth]{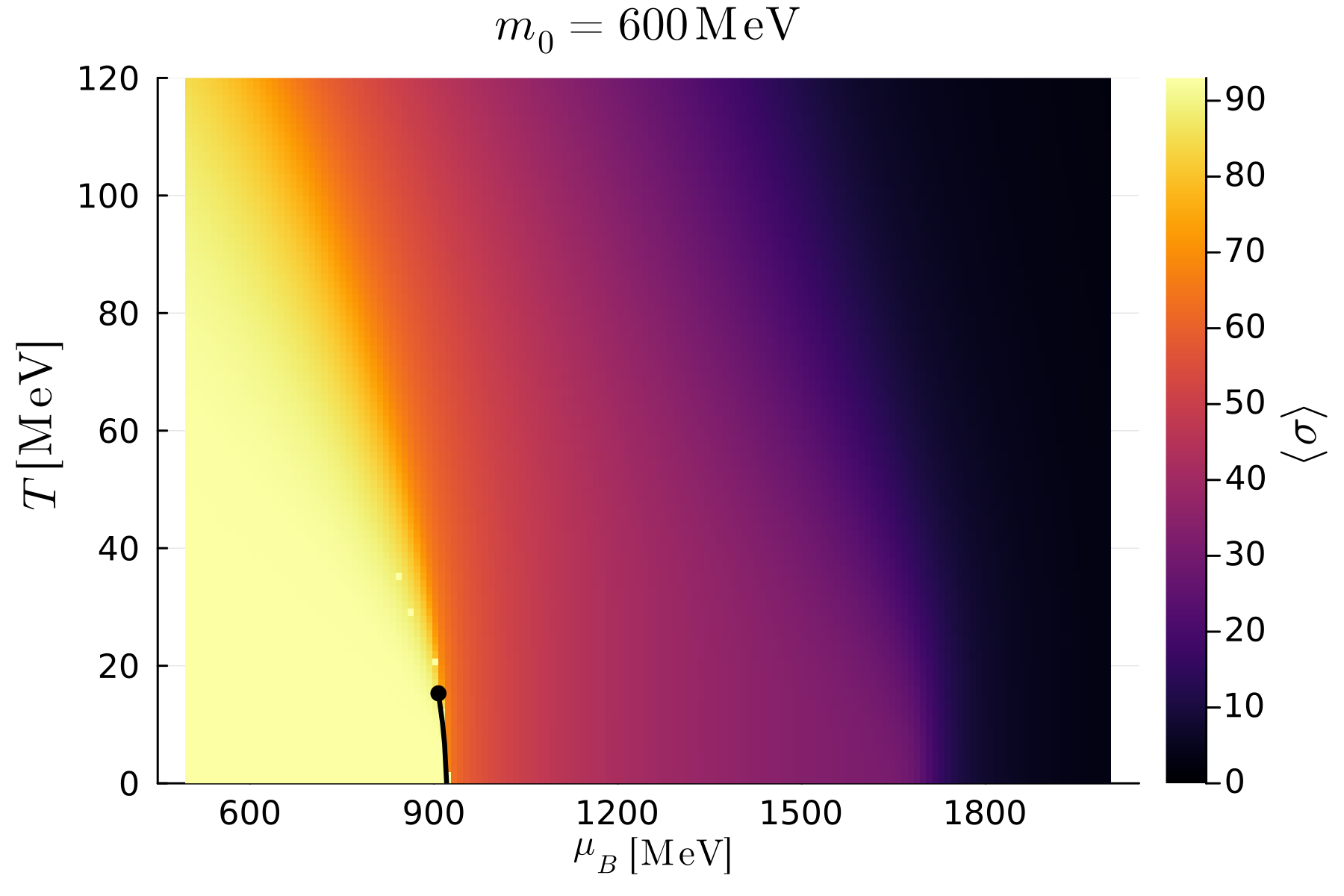}

    \includegraphics[width=0.45\linewidth]{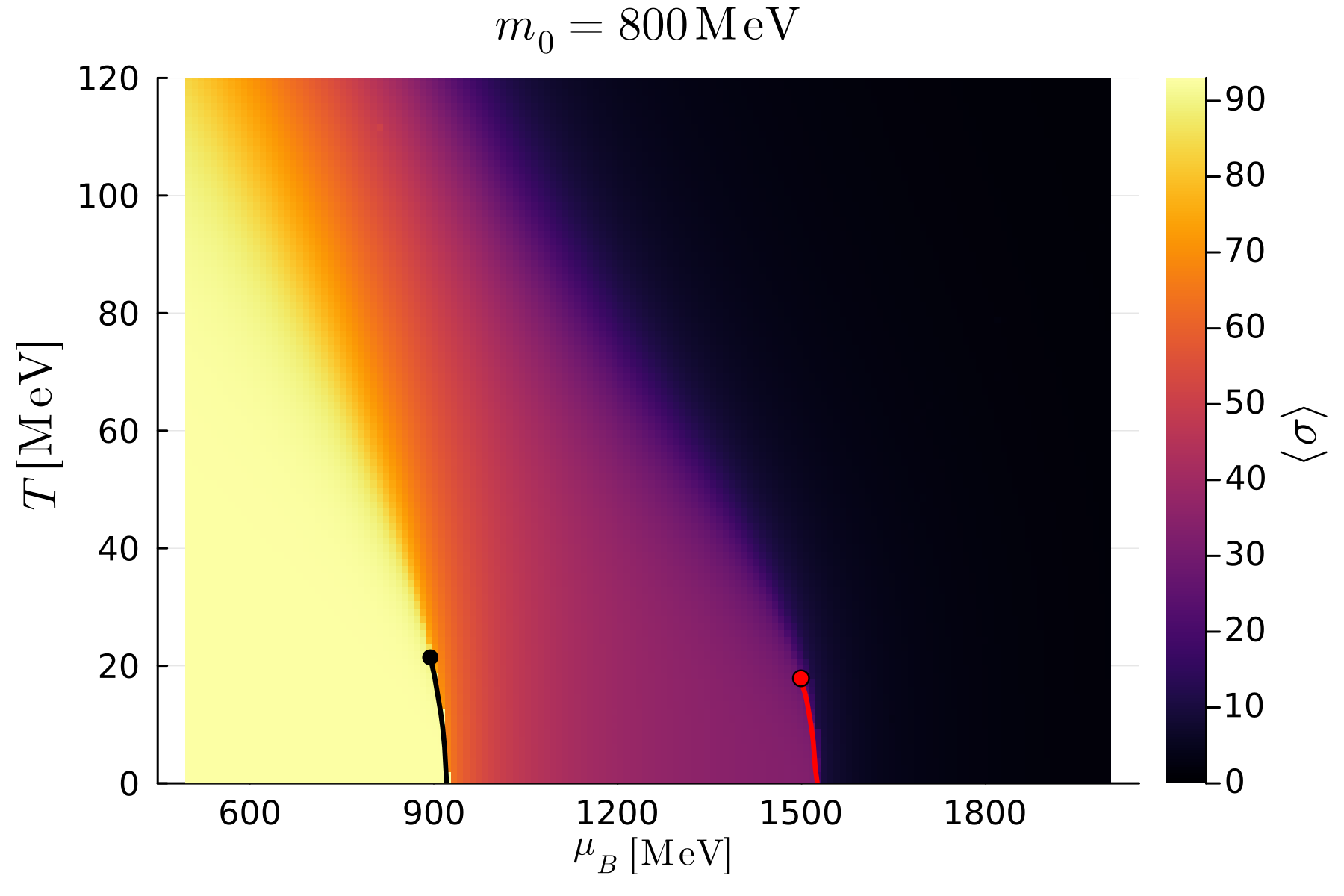}
    \includegraphics[width=0.45\linewidth]{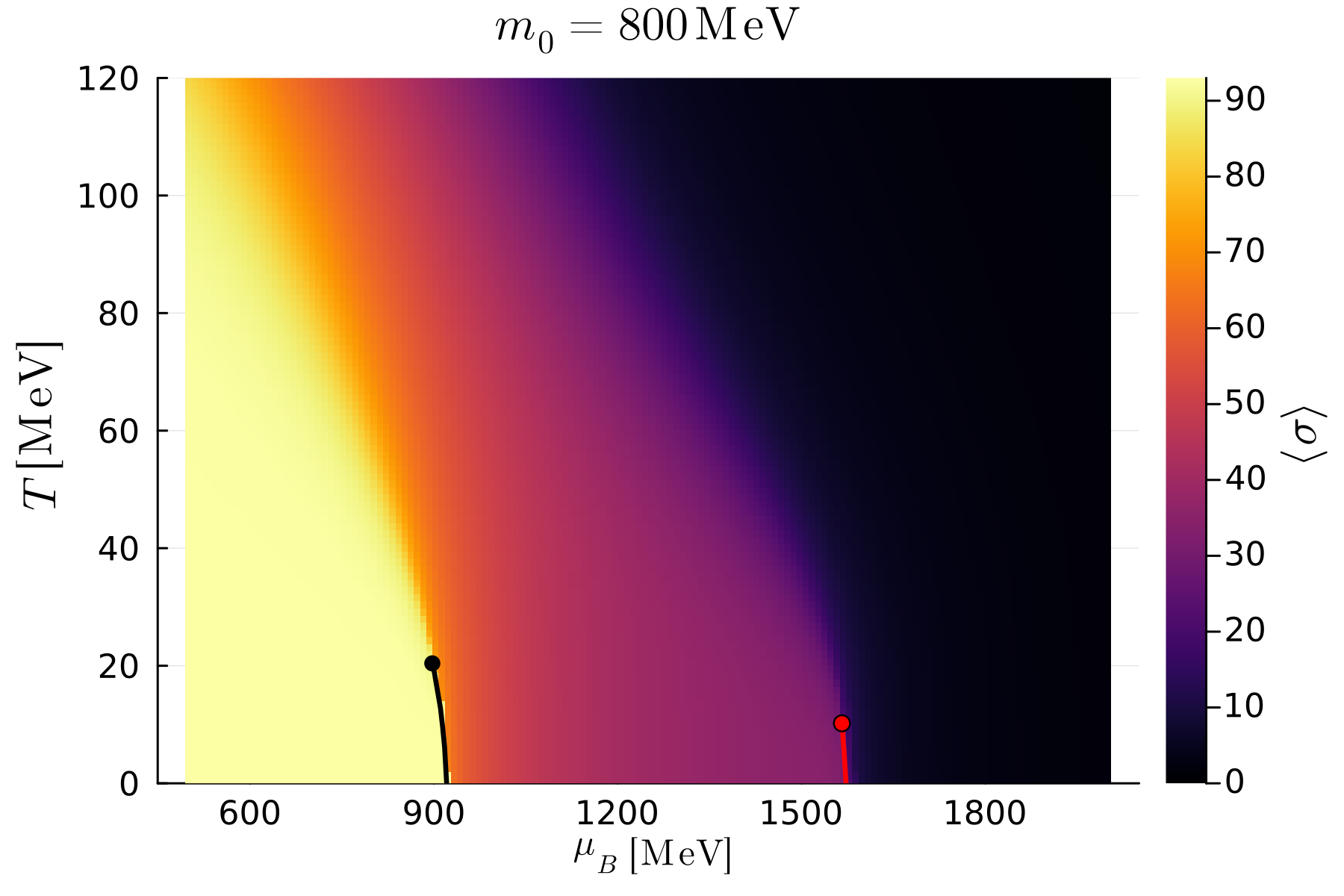}
    \caption{Phase diagrams for symmetric nuclear matter, where the heatmaps display the value of the in-medium chiral condensate 
    $\sigma(T,\mu_B)$ in the $(T,\mu_B)$-plane. The left column does not include the VC while the right one does. The black lines indicate the LG transition while the red lines mark the chiral transition.}

    \label{fig:sym_matt_PhDiag}
\end{figure*}

The temperature $T_c$ of the critical point of the LG phase transition and the symmetry-energy slope parameter,
\begin{equation}
    L=3n_0\pdv{E_\mathrm{sym}}{n}\bigg|_{n_0} \c
\end{equation}
were not included in the fit procedure, but are listed as predictions of the model in the last two columns of Tab.~\ref{tab:fit_quantities} for reference.
The current empirical value for $T_c$ is around 17~MeV but the experimental determination of this quantity is quite challenging and the uncertainty is more than 10\,\% \cite{Karnaukhov:2013gia}. As shown in the table, values of $m_0$ between 600~MeV and 800~MeV are favored by this quantity. The slope parameter $L$ is also marked by relatively large uncertainties, with an accepted empirical value around $60 \pm 10$~MeV \cite{Agrawal:2012pq, Agrawal:2013hha}. More recent studies based on astrophysical observations are consistent within present errors but tend towards somewhat lower values  \cite{Gil:2020wqs,Li:2021thg}. The comparatively large values obtained in the present study (which are in line with Ref.~\cite{Eser:2024xil}), on the other hand, might indicate the  need to include a $\rho - \omega$ interaction term \cite{Fattoyev:2010mx} in the future.

\section{results}\label{sec:Results}

\subsection{Phase Structure}

\begin{figure*}
    \centering
    \includegraphics[width=0.49\linewidth]{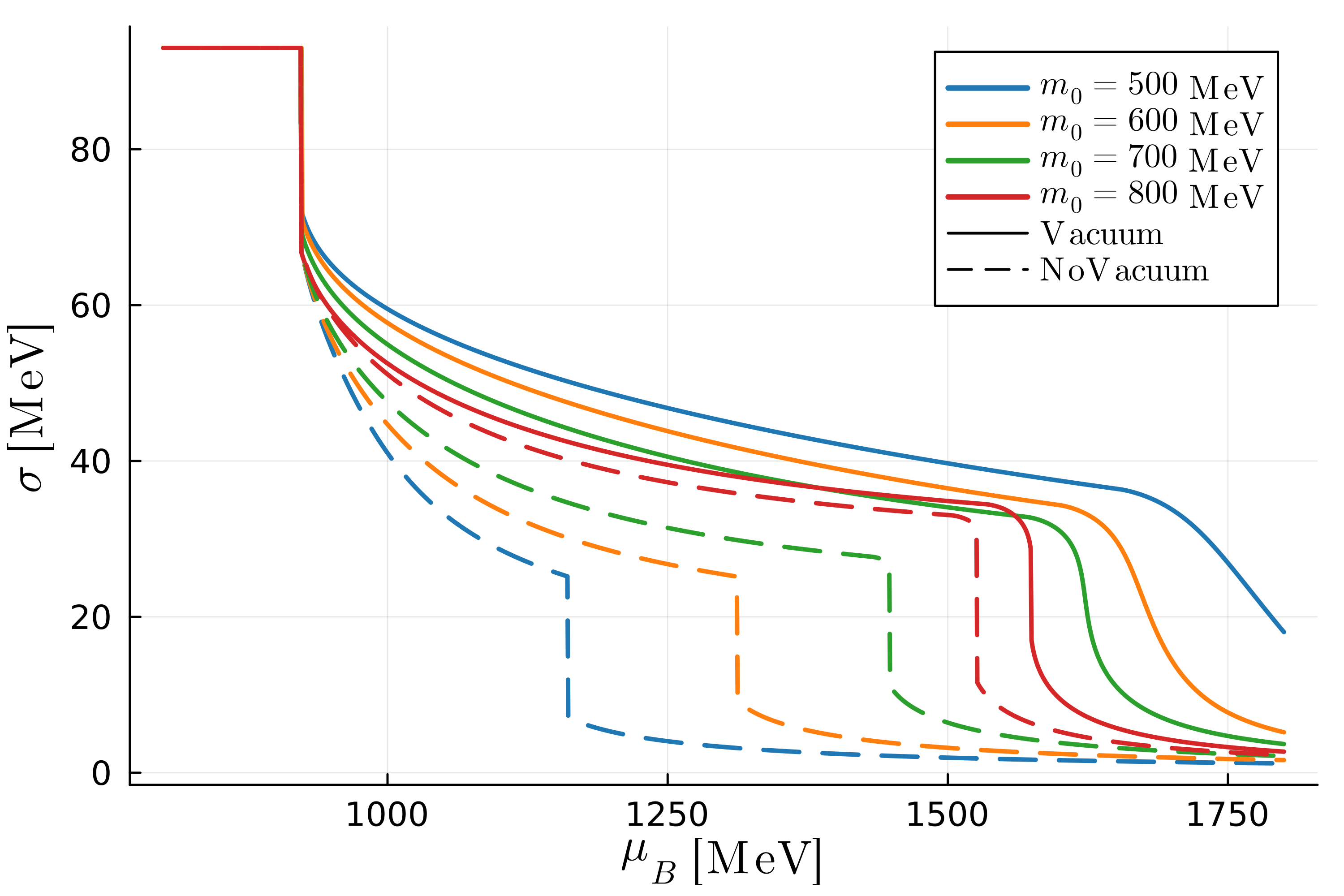}
    \includegraphics[width=0.49\linewidth]{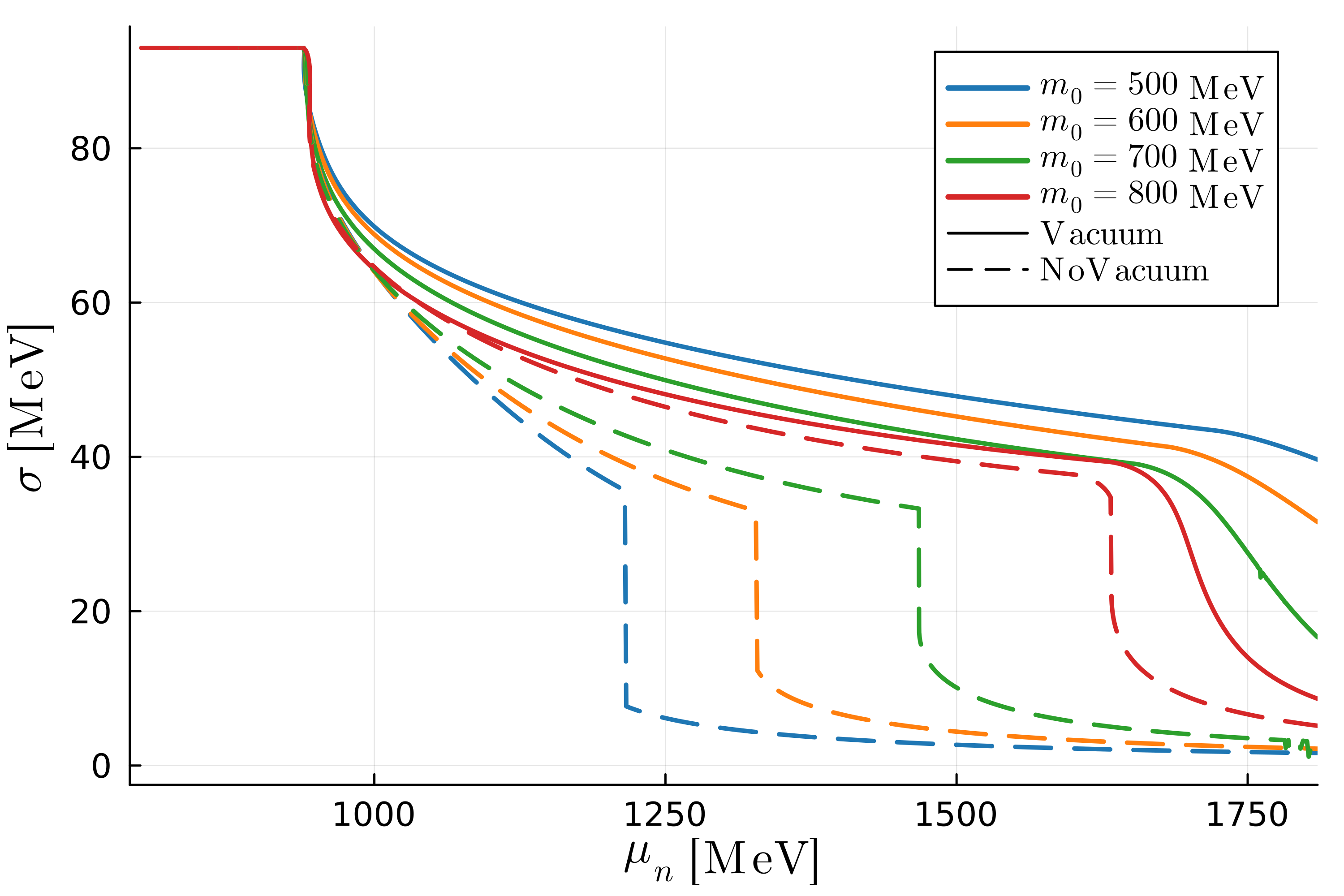}
 \caption{The chiral condensate at $T=0$ for symmetric nuclear matter as a function of $\mu_B=(\mu_p+\mu_n)/2$ (left panel) and pure neutron matter as a function of the neutron chemical potential $\mu_n = \mu_B +\mu_I $ (right panel) for different values of $m_0$. The full lines denote the results with inclusion of the VC while for the dashed lines we leave out the VC. In the left panel, the jump from the vacuum value of $\sigma=f_\pi$ at $\mu_B= 
 m_N -E_B$ represents the onset of self-bound nuclear matter.} 
    \label{fig:chiral_condensate}
\end{figure*}

In Fig. \ref{fig:sym_matt_PhDiag} we display the phase diagram of the model for different values of $m_0$, without (left column) and with (right column) the VC. We report the values of the in-medium chiral condensate as a function of temperature and baryon chemical potential for $m_0=500$~MeV, $600$~MeV and $800$~MeV, respectively. Without the VC we obtain a first-order chiral transition in all cases (red lines). For $m_0=500$~MeV the first-order line extends to $\mu_B=0$.\footnote{We note that the parameter set associated to $m_0=500$~MeV and no VC causes sizable difficulties in the numerical solutions. It is conceivable that forcing the system to reproduce the observables at low values of $m_0$ drives the system into a non-physical minimum of the residual function in Eq.~\eqref{eq:res}.} For the higher $m_0$ values the temperature of the chiral critical point is around that of the LG transition or even below (e.g., around 15~MeV for $m_0=800$~MeV).

\begin{figure}[b]
    \centering
    \includegraphics[width=1\linewidth]{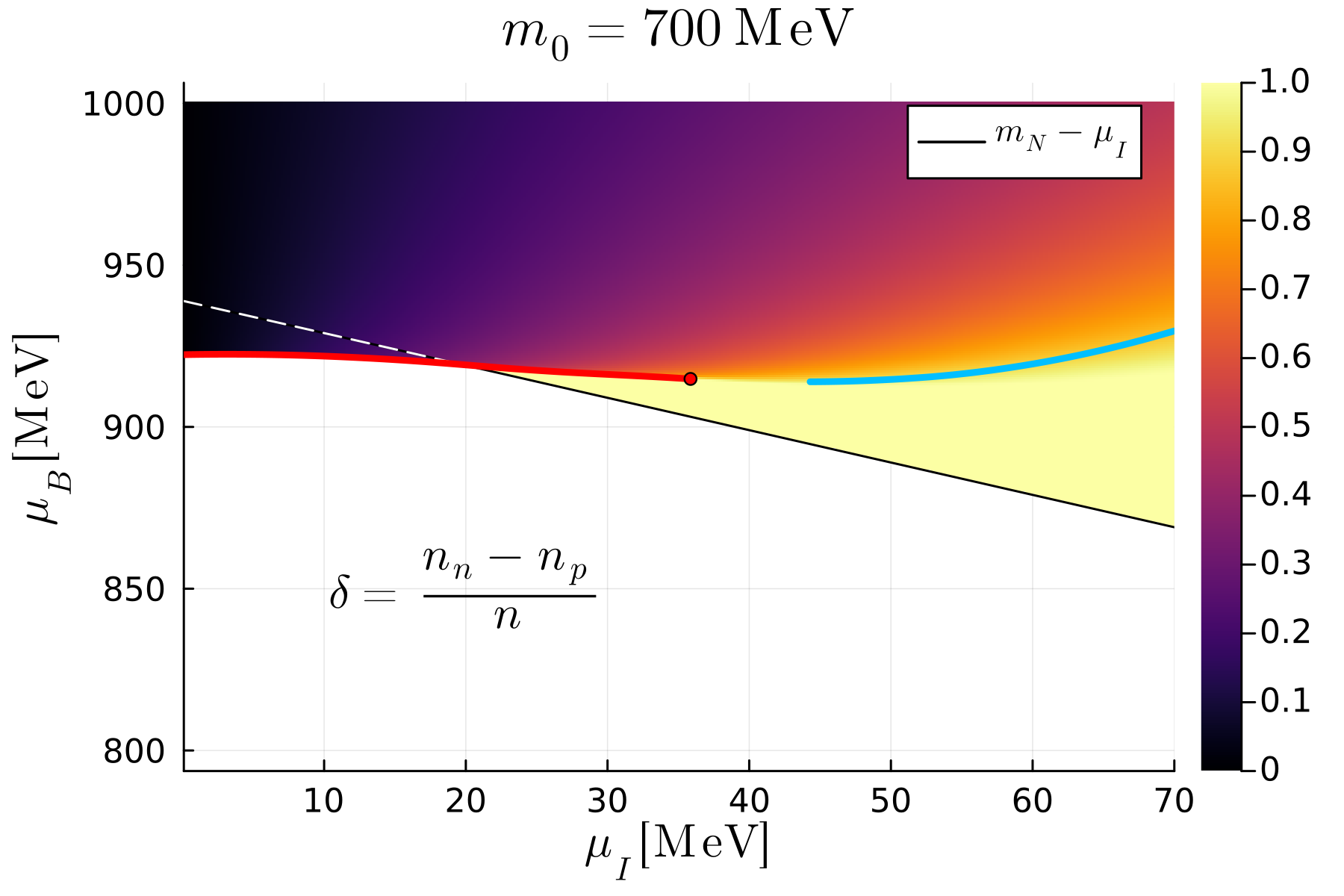}
    \caption{The zero-temperature phase diagram in the $(\mu_I,\mu_B)$-plane. The heatmap displays the value of the isospin asymmetry coefficient $\delta=(n_n-n_p)/n$. The red line marks the discontinuous transition. The diagonal black line separates the vacuum from the onset of unbound neutron matter. The $\beta$-equilibrium line is highlighted in light blue.}

    \label{fig:beta_track}
\end{figure}

\begin{figure*}
    \centering
    \includegraphics[width=0.45\linewidth]{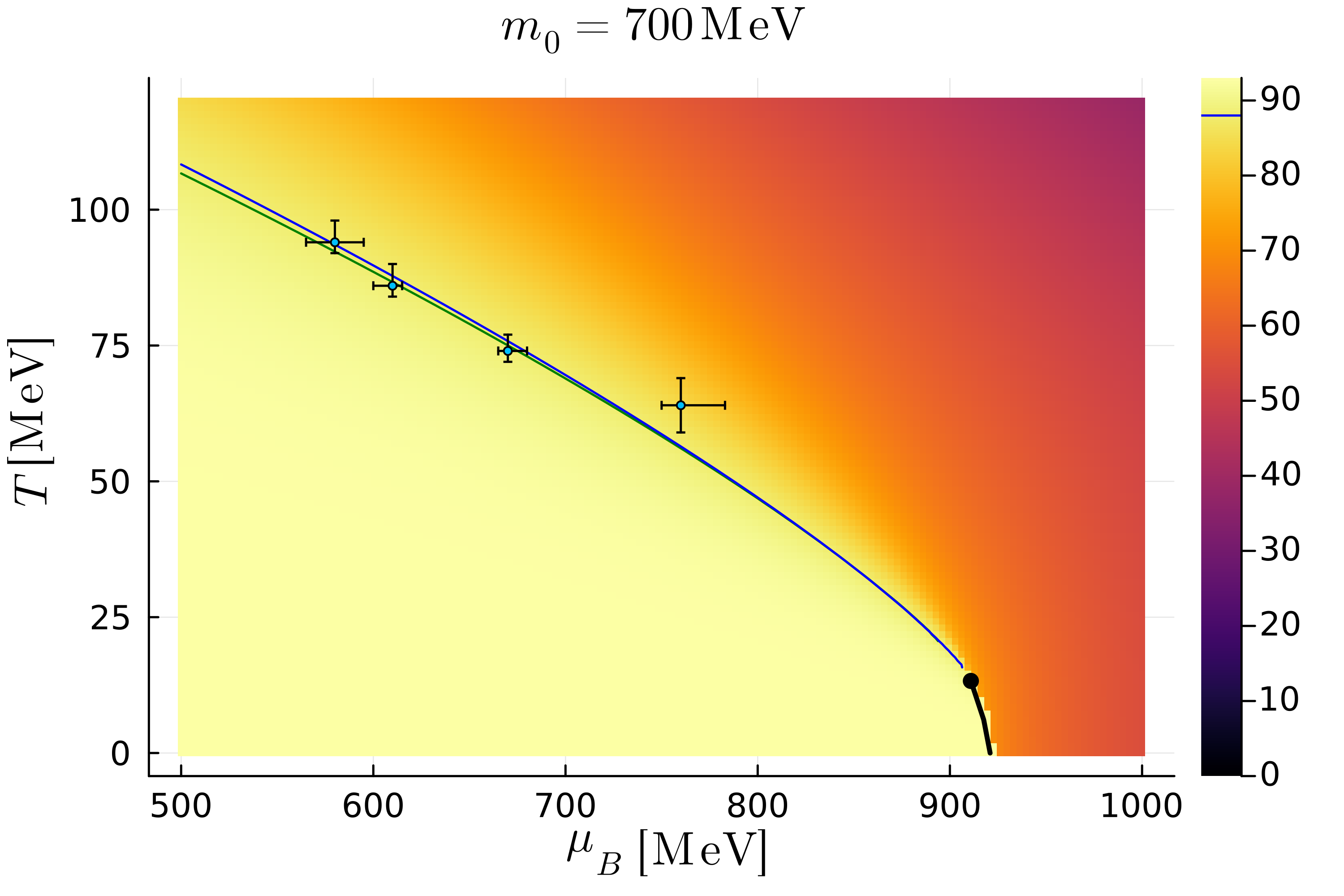}
    \includegraphics[width=0.45\linewidth]{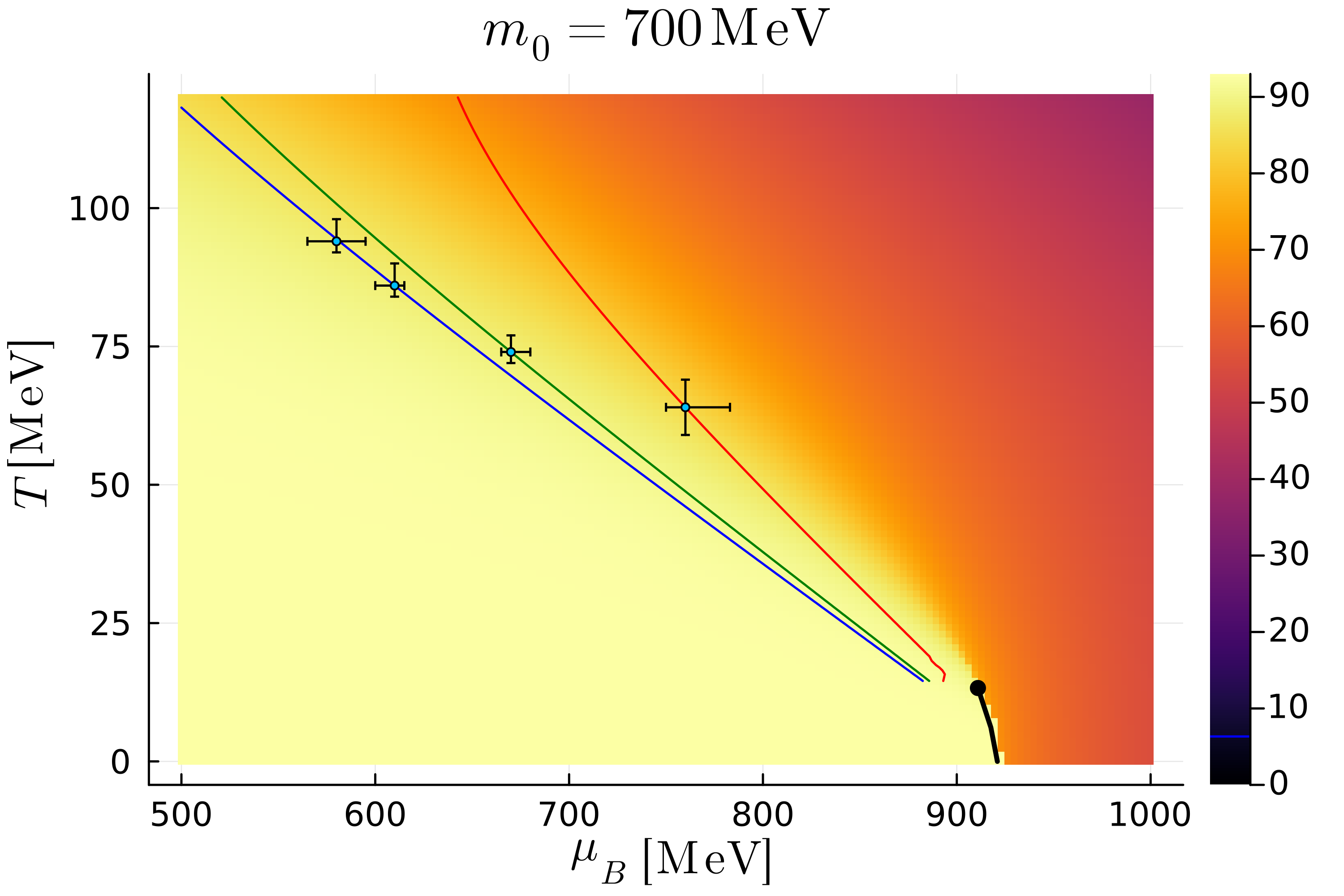}
    \caption{Left panel: zoom-in of the $(\mu_B,T)$ phase diagram (for $m_0=700$~MeV and $\mu_I=0$) around the LG transition. The heatmap displays values of the chiral condensate $\sigma(T,\mu_B)$. The green line represents the chemical freeze-out line at density $n=0.15\,n_0$ as discussed in Ref.~\cite{Floerchinger:2012xd} while the blue line denotes the line of a constant chiral condensate value of $\sigma = 88$~MeV. Right panel: lines of constant entropy/baryon, $S/N $ of about $5.0$ (red), $6.1$ (green) and $6.3$ (blue), passing through the chemical freeze-out points extracted in Ref.~\cite{Andronic:2017pug} \cite{Andronic:2018qqt}. Each line is associated to a different center-of-mass collision energy $\sqrt{s}$: red to 2.7 GeV, green to 3.3 GeV, and blue which passes through the points associated to both 3.8 GeV and 4.3 GeV.}
\label{fig:masses}
\end{figure*}

With the inclusion of the VC the picture changes drastically, especially for the smaller values of $m_0$. The transition is pushed to much higher chemical potentials and becomes a smooth crossover for most values of $m_0$ we have considered, including $m_0 = 700$~MeV (not shown). Only for the largest $m_0=800$~MeV, where the effect of the VC is comparatively small, the chiral first-order transition persists, although at a somewhat higher $\mu_B$ and with a somewhat smaller critical temperature than without the VC. 
Overall, the results we obtain for $m_0=800$~MeV are consistent with those of Refs.~\cite{Eser:2023oii,Eser:2024xil}.

Currently, there is no experimentally conclusive direct evidence for a first-order chiral phase transition in the QCD phase diagram although the existence of a chiral critical point is expected from state-of-the-art QCD calculations with functional methods \cite{Fu:2019hdw,Gunkel:2021oya} and a majority of QCD-based models.
Since in our study the chiral transition is located in very high-density regions of the phase diagrams (at densities $n_\chi \sim 10 \, n_0$), we expect the details of the hard-core repulsion of the nuclear interaction to play a crucial role.\footnote{Implicitly, using the PDM assumes that baryonic degrees of Freedom survive chiral restoration. However, the densities found in the present study could possibly be high enough to make a transition to deconfined quark matter~\cite{Otto:2019zjy} which, of course, cannot be described by a hadronic model such as the PDM.}

In Fig.~\ref{fig:chiral_condensate} we summarize the evolution of the zero-temperature chiral condensate for the $m_0$ values between 500~MeV and 800~MeV, with the parameters of Tab.~\ref{tab:fit_prms}, for symmetric nuclear matter (left panel) and pure neutron matter (right panel). The latter is relevant for evolved isolated NS, to be discussed below. 

For isospin symmetric matter in the left panel the first jump is located at $m_N-E_B \approx 923$~MeV and is associated with the onset of self-bound nuclear matter.\footnote{We do not study the mechanically unstable region (e.g., low-density phase separation) of nuclear matter but only investigate stable homogeneous matter. In particular we solve all equations at fixed $\mu_B$ which means that densities below $n_0$ cannot be explored for symmetric nuclear matter.}
The dashed lines, which do not include VC, show a second jump corresponding to a first-order chiral phase transition. The position of this jump strongly depends on the value of $m_0$ and is close to the first one for small values of $m_0$.
The inclusion of the VC makes the transition a smooth crossover, except for the largest  $m_0=800$~MeV. In this case the effect of the VC is much less dramatic and affects only the position of chiral transition and the size of the jump that is reduced instead of being eliminated. Importantly, however, we note that reducing the chiral baryon mass $m_0$ from 800~MeV down to 500~MeV with or without the VC has the opposite effect on the position of the chiral transition, where the $m_0$ dependence is stronger without the VC. In particular, a low-density chiral transition is only possible for small chiral baryon mass $m_0 $ in combination with the no-sea approximation, i.e.~without the VC. We therefore conclude that it is an artifact of this rather crude approximation entirely neglecting the VC.

In the right panel of Fig.~\ref{fig:chiral_condensate} there is no first jump but a smooth onset of repulsive pure neutron matter placed at $\mu_n = m_N$, where $\mu_n = \mu_B + \mu_I$ is the chemical potential for neutrons with our conventions for $\mu_I $ (as introducing an excess of neutrons). The conclusions on the $m_0$ dependence of the chiral transition are qualitatively the same as for symmetric nuclear matter.

Fig.~\ref{fig:beta_track} shows the value of the isospin asymmetry parameter $\delta=(n_n-n_p)/n$ as a function of $\mu_B$ and $\mu_I$ at vanishing temperature, i.e.~the zero-temperature phase diagram. We choose $\delta$ because it highlights the various phases which are characterized by different neutron and proton fractions. The phase diagram is displayed up to $\mu_I = 70~\mbox{MeV} \approx m_\pi/2$ because at isospin chemical potential $\mu_I = m_\pi/2$ one expects the onset of pion condensation that we do not consider in this work but refer to the analogous discussion in the quark-meson model \cite{Kamikado:2012bt,Brandt:2025tkg}. 
Negative values of $\mu_I$ produce the same zero-temperature phase diagram with $\delta \to -\delta$. We can therefore limit our discussion to $\mu_I \geq 0$ (here for neutron excess) without loss.

The lower, empty part of the figure, represents the zero-density (Silver-Blaze) region associated with the vacuum of the theory. Following the $\mu_B$ axis at $\mu_I=0$ one sees the onset of self-bound  symmetric nuclear matter  at $\mu_B =  m_N-E_B \approx 923$~MeV. The step between the white vacuum region and the dark symmetric matter region ($\delta = 1$) is discontinuous in density with a jump from $n=0$  to $n_0 \approx 0.16$~fm$^{-3}$. For non-vanishing but small $0 < \mu_I\lesssim 20$~MeV the red line indicated in Fig.~\ref{fig:beta_track} separates predominantly symmetric nuclear matter (with a small isospin asymmetry) discontinuously from the vacuum.
At $\mu_I \approx 20 $~MeV, the red line intersects with the black line along $\mu_B = m_N - \mu_I$ which marks the energy cost to produce a single unbound neutron (neutron drip). Beyond this point the red line separates predominantly symmetric nuclear matter from essentially pure neutron matter. The separation is discontinuous in density up to the red critical point at around $\mu_I \approx 36 $~MeV where the discontinuity vanishes in a continuous second-order transition. Beyond this critical point in the zero-temperature phase diagram the 
LG transition turns into a smooth crossover. This critical point is the limit $T_c(\mu_I) \to 0 $ of the LG critical point starting at  $T_c \approx 17$~MeV for symmetric nuclear matter at $\mu_I = 0$. 


The bright yellow region above the $\mu_B = m_N - \mu_I$ line  is the phase of pure neutron matter ($\delta = 1$). Since neutron matter is not self-bound, the transition between vacuum and the yellow region, across the black line, is continuous in density. The top part of the diagram, above the yellow region, is characterized by arbitrary mixtures of neutrons and protons in asymmetric nuclear matter.
The light-blue curve indicates mixtures of neutrons and protons that satisfy the conditions of $\beta$-equilibrium in neutron star matter, to be discussed in the next section.

The qualitative aspects of the zero-temperature phase diagram shown in Fig.~\ref{fig:beta_track} are model-independent features of QCD. As such they are reproduced by the PDM, here exemplified with $m_0 = 700$~MeV, but essentially unchanged with the other parameter sets of Tab.~\ref{tab:fit_prms}, all fitted to the saturation properties of nuclear matter, as well.  
The omission or inclusion of the VC is not as important as it is for Fig \ref{fig:sym_matt_PhDiag} either, here. The only noticeable effect is a slight (few MeV) shift in the position of the critical point which is, however, within the uncertainties of the phenomenological properties of nuclear matter, i.e.\ mainly due the variations in $\sigma(n_0) $, $L$ and $T_c$, cf.~Tab.~\ref{tab:fit_quantities}. 

 The evolution of the in-medium chiral condensate in the vicinity of the nuclear-matter LG transition may have interesting consequences for heavy-ion collisions at beam energies of the HADES and future CBM experiments at FAIR/GSI. As seen in the left panel of Fig.~\ref{fig:chiral_condensate} the $T=0$ chiral condensate jumps discontinuously at the onset of self-bound nuclear matter with saturation density $n_0$. In the $(\mu_B,T)$ phase diagram this discontinuity persists up to the critical point of the LG transition, from where on it changes into a rather sharp crossover, which only gradually widens with decreasing $\mu_B$, as seen in Fig.~\ref{fig:masses}.
 
The chemical freeze-out points, extracted from Refs.~\cite{Andronic:2017pug,Andronic:2018qqt}, are located along this crossover region, and are well described by a  line of constant density with $n=0.15\,n_0$ (green line in the left panel of Fig.~\ref{fig:masses}) as previously noted in Ref.~\cite{Floerchinger:2012xd}. Interestingly, they are also described by a line of constant chiral condensate with a value of $\sigma=88$~MeV (blue line in the left panel of Fig.~\ref{fig:masses}). 
The fact that the contour lines of the density and the chiral condensate at their respective values emerging from the critical point stay so close to one another is an indication of a rather sharp crossover in this region of the phase diagram which also intuitively explains the proximity of the freeze-out points. This coincidence of the contour lines will not persist towards much smaller values of $\mu_B$.

In the PDM, sizable modifications of $\sigma(
T,\mu_B)$ directly translate into a reduction of the vacuum mass splitting, $\Delta m_\pm(f_\pi) = 571$~MeV, between the $N(939)$ and the $N^*(1535)$. For $T=0 $ and $\sigma(n_0) \simeq 68$~MeV at saturation, for example, one obtains $\Delta m_\pm \simeq 422$~MeV, and  $\Delta m_\pm \to 0$ with the gradual restoration of chiral symmetry.
With the value of $\sigma \simeq 88$~MeV in the vicinity of the chemical freeze-out line in the left panel of Fig.~\ref{fig:masses} it already reduces to $\Delta m_\pm \simeq 541 $~MeV and hence falls below the mass of the $\eta $ meson so that the $N^*(1535) \leftrightarrow \eta N$ channel closes.
This may have observable implications for dilepton invariant-mass distributions at HADES/CBM energies \cite{Larionov:2021ycq,Geurts:2022xmk}, dilepton angular distributions \cite{Seck:2023oyt} and the electrical conductivity of nuclear matter \cite{Wambach:2023tba}.        


 \begin{figure}[t]
    \centering
    \includegraphics[width=\linewidth]{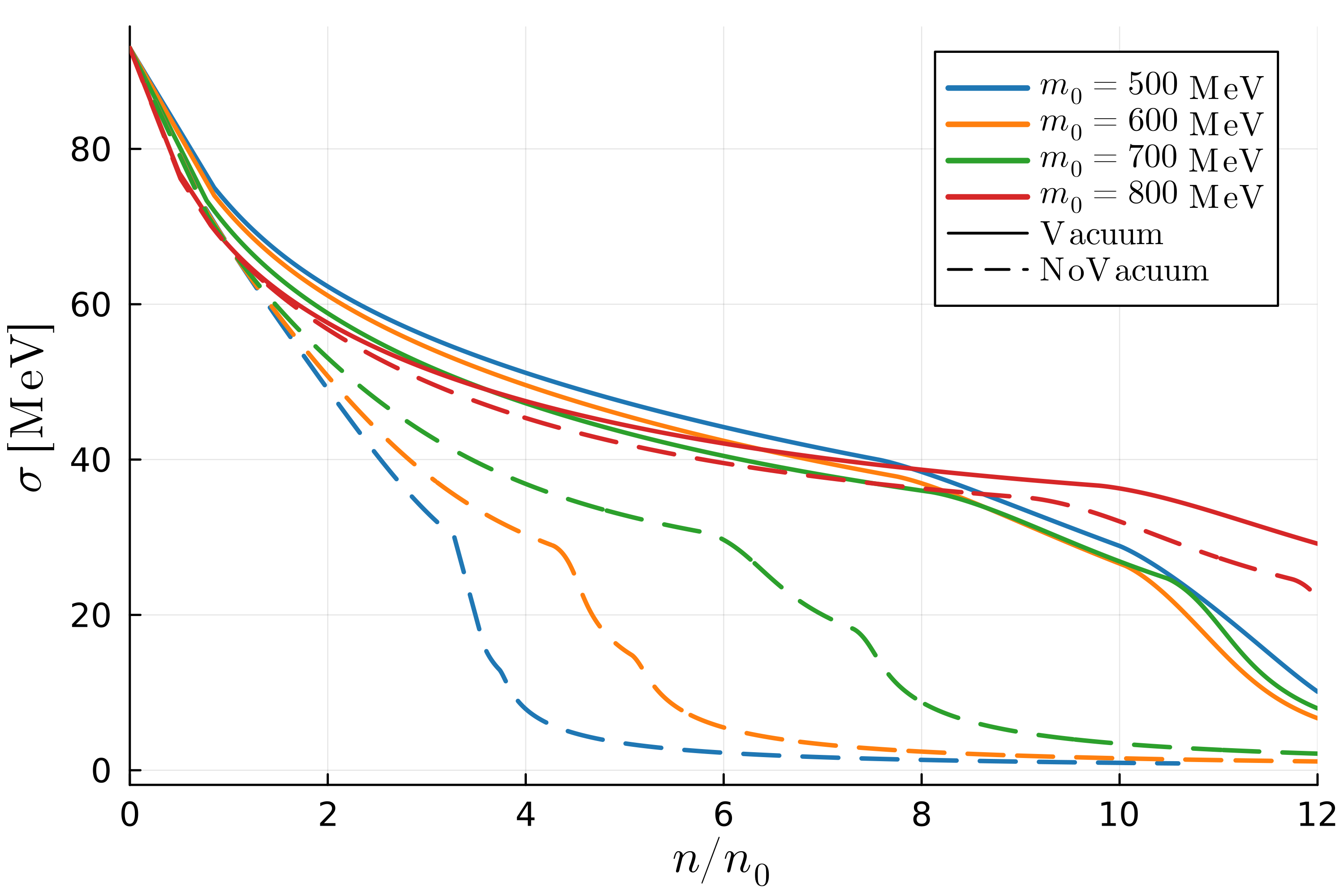}

    \caption{The chiral condensate in $\beta$-equilibrated NS matter at $T=0$ as a function of $n/n_0$ for various $m_0$ values, without (dashed lines) and with the VC (full lines). While for $m_0=800$~MeV the VC only induces small modifications, as before, for the lower values of $m_0$ the behavior of $\sigma$ field changes significantly when the VC is neglected.}
    \label{fig:chiral_condensate_beta}

\end{figure}

\subsection{Beta-Equilibrated Matter and Neutron Stars}

In this section we discuss the implications of the thermodynamics of the PDM for isolated NS and NS mergers, focusing on the role of the in-medium chiral condensate for the EoS.

Since NS are charge neutral and in electro-weak equilibrium, the following conditions have to be imposed  
\begin{align}
    \mu_n-\mu_p &= \mu_e=\mu_\mu \c \\
    n_p&=n_e+n_\mu \c
\end{align}
where the subscripts $p$ and $n$ denote the iso-doublet of protons and neutrons together with their respective parity partners ($p^*$ and $n^*$) while $e$ and $\mu$ denote electrons and muons (it is assumed that the corresponding neutrinos, $\nu_e$ and $\nu_\mu$, freely escape during the NS evolution).

For zero-temperature neutron-star matter these conditions fix a line in the $(\mu_I,\mu_B)$-plane which is represented by the light blue line in Fig.~\ref{fig:beta_track}. It begins in a region of high isospin asymmetry $\delta$, at $\mu_I$ values beyond the critical point of the LG transition and moves continuously through regions of lower $\delta$ due to the appearance of $p$, $e$ and $\mu$ with increasing $\mu_I = (\mu_n-\mu_p)/2 $.\footnote{The phase diagram is calculated fixing a pair $(\mu_I, \mu_B)$ and computing the corresponding $n$ and $n_I$ that minimize the grand potential. This procedure, actually, prevents the exploration of the system at certain baryon and isospin densities. For this reason we report the $\beta$-equilibrium line in Fig.~\ref{fig:beta_track} only where it is compatible with the underlying phase diagram.}


In Fig. \ref{fig:chiral_condensate_beta} we display the chiral condensate as a function of baryon density in $\beta$-equilibrated NS matter. 
The chiral condensate evolves in the low density part as expected for pure neutron matter. The main difference is in the chiral transition that is triggered by the onset of the parity partner of the neutron, the $n^*$, at first. With increasing $p^*$ fraction, eventually another drop is visible leading to a splitting of the chiral crossover. The splitting structure is more visible without the VC. It is washed out with the VC leaving a soft crossover that extends over a wide range of densities (roughly between $8 \, n_0$ and $10 \, n_0$) and makes the evolution of the condensate less sensitive to $m_0$ (at least between $500$~MeV and $ 700$~MeV).  

After the abundant evidence of the importance of the VC, we focus exclusively on  the parameter sets with the VC included, in the discussion of the cold NS properties from the PDM for the rest of this section.


\begin{figure*}
    \includegraphics[width=0.49\linewidth]{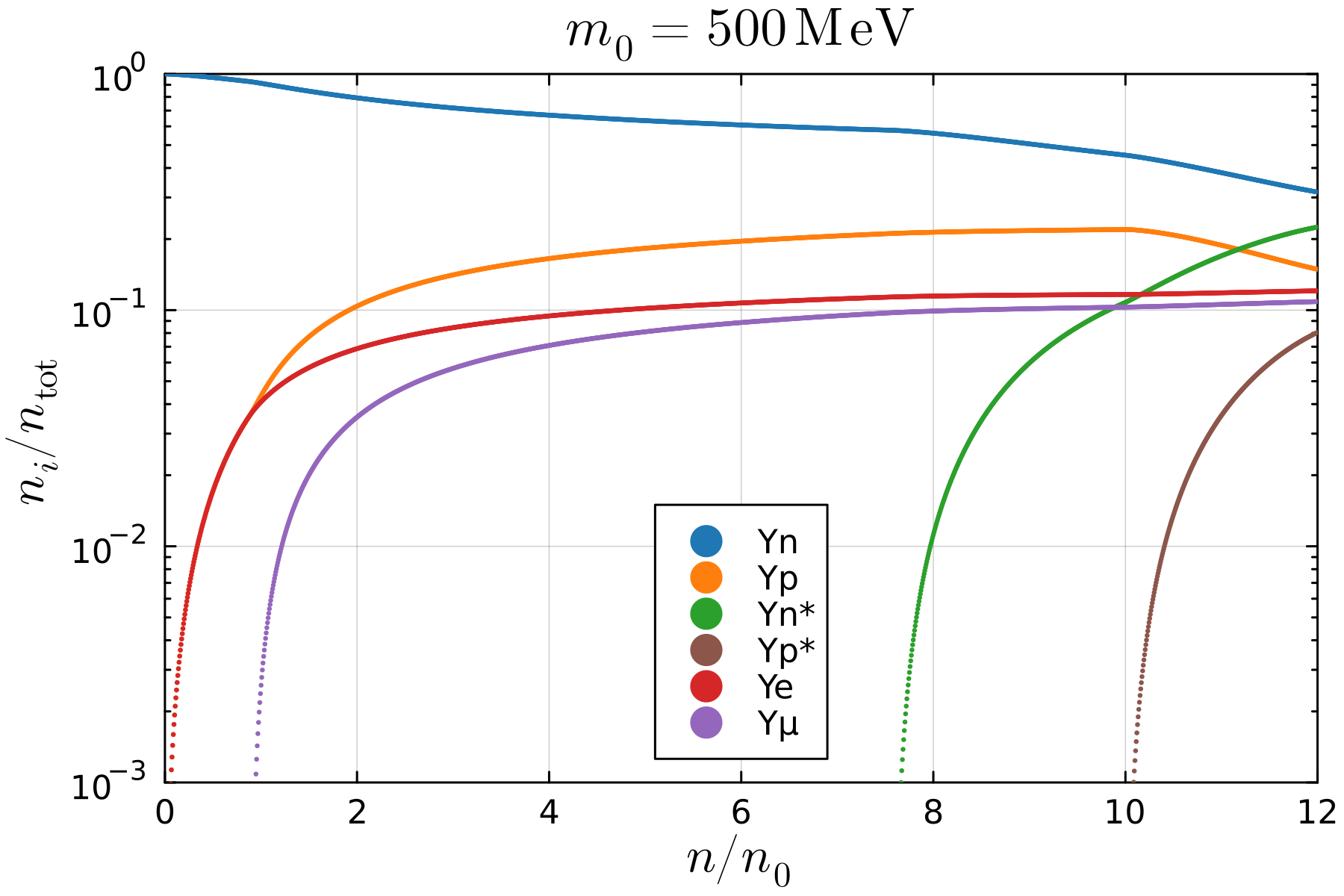}
    \
    \includegraphics[width=0.49\linewidth]{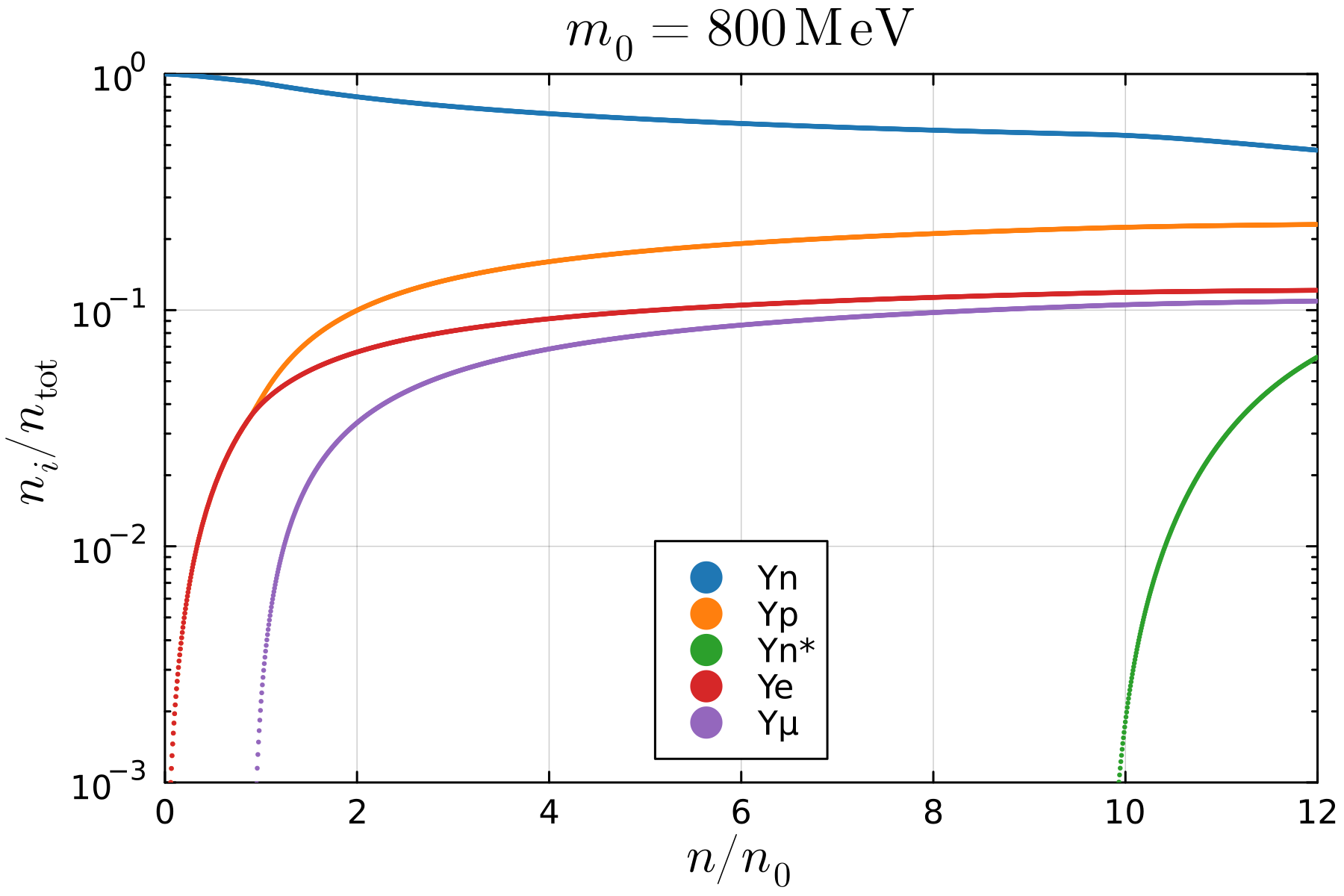}
    \caption{Relative abundances of the various species in cold $\beta$-equilibrated NS matter as a function of baryon density $n/n_0$.}
    \label{fig:abundances}
\end{figure*}

\begin{figure*}[ht]
    \includegraphics[width=0.49\linewidth]{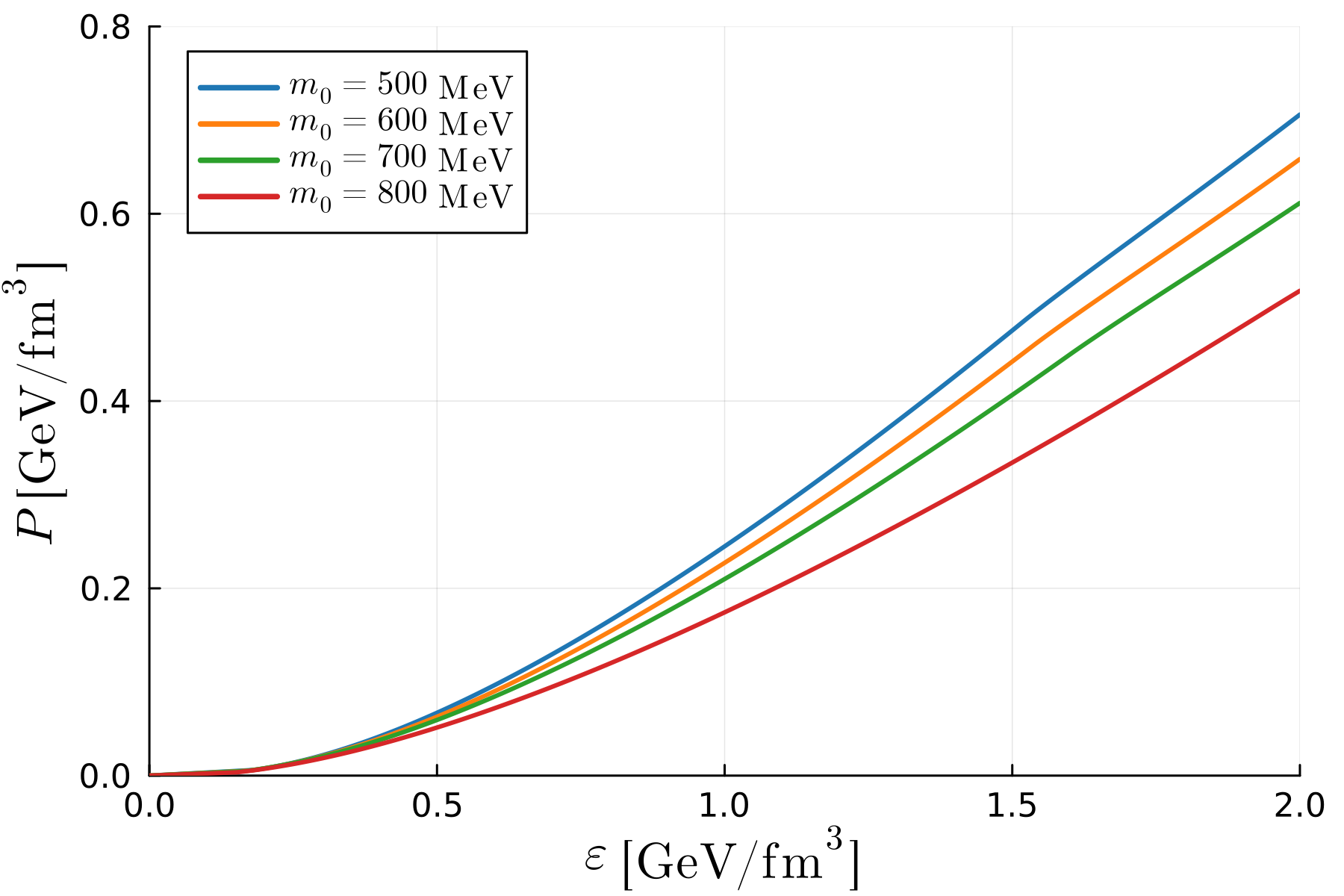}
    \includegraphics[width=0.49\linewidth]{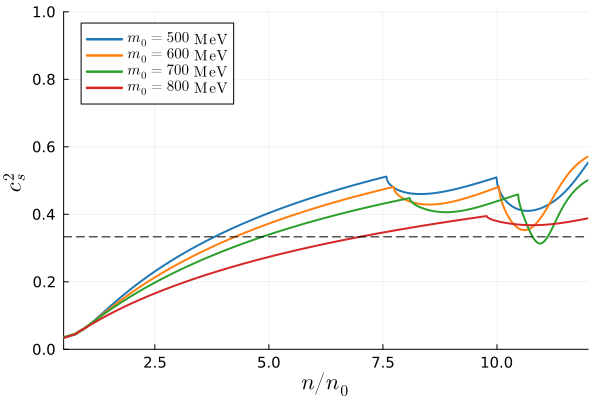}
    \caption{The EoS (left panel) and the squared speed of sound (right panel) of cold NS matter for the various values of $m_0$. The dashed line in the right panel denotes the conformal limit $c_s^2 = 1/3$. }
    \label{fig:sound_eos}
\end{figure*}

In Fig. \ref{fig:abundances} the relative abundances of neutrons and protons as well as their parity partners, $n^*$ and $p^*$, together with electrons and muons are shown for zero-temperature $\beta$-equilibrated matter with $m_0=500$~MeV (left panel) and $m_0=800$~MeV (right panel). While $n$ is the baryon density as before, with $n_\mathrm{tot}$ we here denote the total particle-number density. Since we have assumed proton and neutron isospin degeneracy, there is no threshold for the appearance of electrons.

\begin{figure*}
    \centering
    \includegraphics[width=0.49\linewidth]{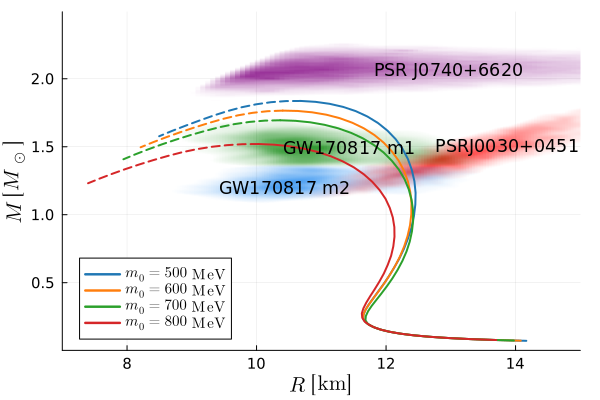}
    \includegraphics[width=0.49\linewidth]{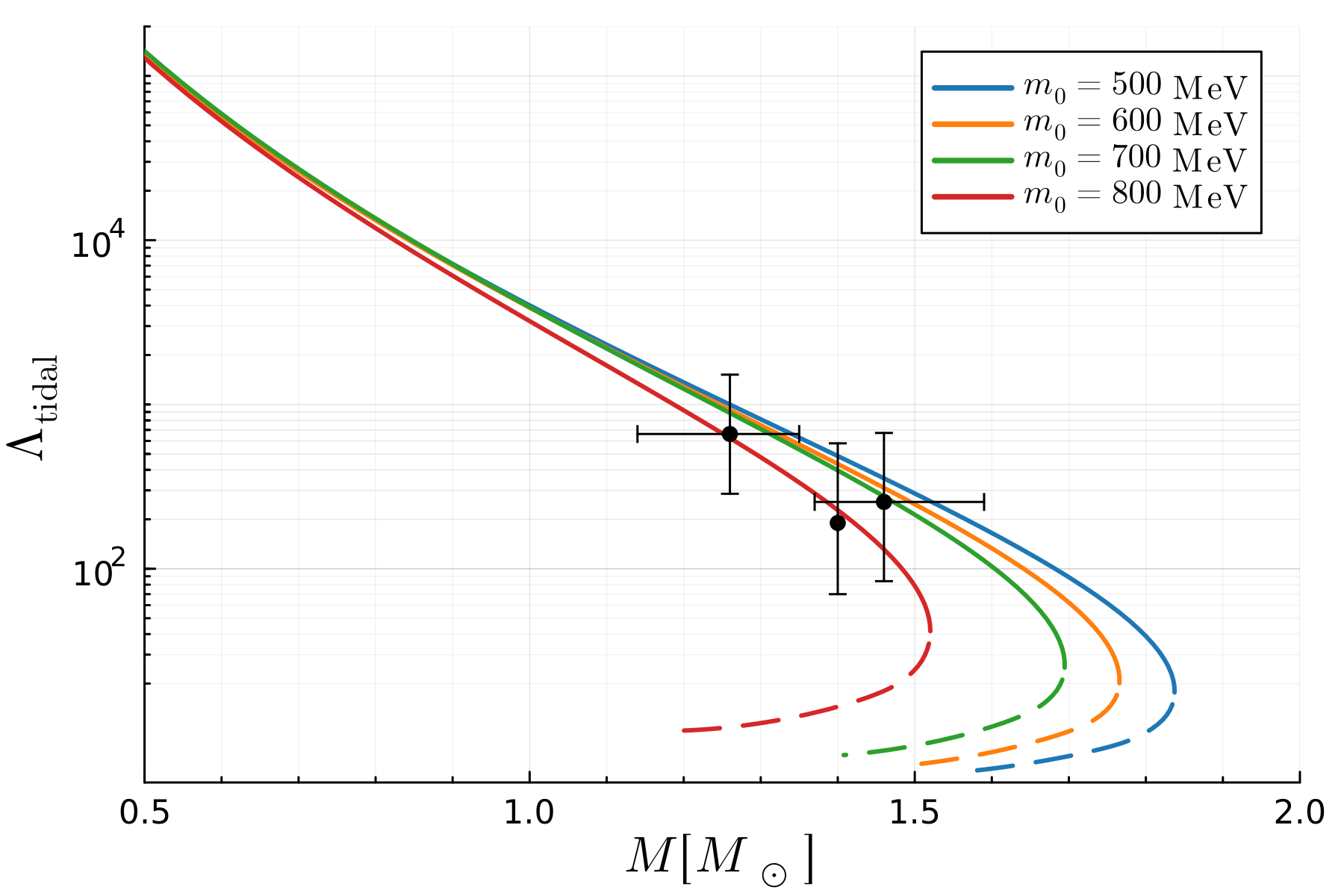}
    \caption{Left panel: The M-R relation for evolved NS, obtained with the parameter sets discussed in the text.
    The shaded colored areas indicate constraints from astrophysical observations given in \cite{Miller:2019cac}, \cite{Dittmann:2024mbo} and \cite{LIGOScientific:2018cki}.
    Right panel: The tidal deformability as a function of the stellar mass for the various values of $m_0$. All cases are consistent with the 90\% CL observation value of GW170817 (data points) \cite{LIGOScientific:2018cki}\cite{Fasano:2019zwm}.
    In both panel the dashed lines highlight the unstable branch of the curves.}
    \label{fig:M-R}
\end{figure*}

Finally, we discuss the EoS of cold NS matter. Fig.~\ref{fig:sound_eos} displays the pressure $P$ as a function of energy density $\epsilon$ (left panel), and the squared speed of sound $c_s^2$ as function of density (right panel). In $c_s^2$ the effects of the chiral transition manifest themselves as cusps and troughs reflecting the onset of the $n^*$ and $p^*$ degrees of freedom.

The speed of sound exceeds the conformal limit of asymptotically free quark matter. This is not unexpected 
for an effective baryonic model with short-range vector repulsion, however. In fact, with $\bar\omega \propto n $ for large $\mu_B $ the energy associated with the $\omega $ expectation value is not negligible compared to $\mu_B$, and one necessarily obtains $c_s^2\to 1$ while the conformal limit requires the chemical potential to be the only relevant energy scale for $n \to \infty$. In principle, the expected conformal limit with $c_s^2\to 1/3 $ could be restored by adding quartic self-interactions $\propto \omega^4$ \cite{Mueller:1996pm,Fraga:2022yls}. We do not apply this fix here because the short-distance repulsion between the baryons as an effective QCD interaction is eventually expected to be dissolved with asymptotic freedom in the limit of large $\mu_B$. For a recent discussion of the speed of sound and its approach to the conformal limit see Ref.~\cite{Fukushima:2024gmp}.

Comparing with the values in Tab.~\ref{tab:NS} we can see that for each $m_0$ the chiral transition happens after the maximum density supported by the neutron star. This implies that in the two-flavor PDM we encounter no chirally restored matter in cold neutron stars once the VC is taken into account. The possibility of chirally restored matter in proto-neutron stars or in NS post-mergers remains open.

\begin{table}[t]
     \centering
    \textbf{No Vacuum Contribution}\\
    \vspace{1 mm}
    \begin{tabular}{|c|c|c|c|}
        \hline
        $m_0$ [MeV] & $M_\mathrm{max}$ [$M_\odot$ ] & $\epsilon_\mathrm{max}$ [MeV/fm$^3$] & $n_\mathrm{max}$ [$n_0$] \\
        \hline
        500 & 1.94 & 1479 & 7.42 \\ 
        600 & 1.87 & 1395 & 7.19 \\ 
        700 & 1.76 & 1383 & 7.22 \\ 
        800 & 1.53 & 1576 & 8.26 \\ \hline
        \end{tabular}\\
    \vspace{1 mm}
    \textbf{Vacuum Contribution}\\
    \vspace{1 mm}
    \begin{tabular}{|c|c|c|c|}
    \hline
        $m_0$ [MeV] & $M_\mathrm{max}$ [$M_\odot$ ] & $\epsilon_\mathrm{max}$ [MeV/fm$^3$] & $n_\mathrm{max}$ [$n_0$] \\
        \hline
        500 & 1.84 & 1383 & 7.05 \\ 
        600 & 1.77 & 1527 & 7.7 \\ 
        700 & 1.69 & 1525 & 7.79 \\ 
        800 & 1.52 & 1617 & 8.43 \\  
        \hline
    \end{tabular}
    \caption{Maximum mass, as well as the energy-density and the number density in the NS center for the various values of $m_0$ considered here, without (top) and with the VC (bottom) for comparison.}
    \label{tab:NS}
\end{table}
 
Fig.~\ref{fig:M-R} reports, in the left panel, the M-R relations resulting from our EoS's matched to the HS(DD2) EoS~\cite{HSDD2, Typel:2009sy} for the low density part up to $n_0$. The solution of the TOV equation is provided by the software toolkit \texttt{TOVExtravaganza} \cite{TOVExtravaganza}.
As we can see from  Fig.~\ref{fig:M-R} and Tab.~\ref{tab:NS}, the maximum masses are below 2 solar masses for each parameter set. This implies that the present model is not able to reproduce the astrophysical constraint by the mass of PSR J0740+6620 \cite{Dittmann:2024mbo}, with the smaller values of $m_0$ coming closer to this constraint than the larger ones.  The right panel of Fig.~\ref{fig:M-R} shows the dimensionless tidal deformability $\Lambda_\mathrm{tidal}$ as a function of the stellar mass. Here,  we find agreement with current gravitational-wave observations \cite{LIGOScientific:2018cki} for all values of $m_0$.

Another key quantity of interest for NS matter, derived from the EoS, is the thermal index,
\begin{equation}
    \Gamma_\mathrm{th} = 1+ \frac{P_\mathrm{th}}{\varepsilon_\mathrm{th}} \c
\end{equation}
with
\begin{align}
    P_\mathrm{th} &= P(\mu_B,\delta,T)-P(\mu_B,\delta,0) \c \\
    \varepsilon_\mathrm{th} &= \varepsilon(\mu_B,\delta,T)-\varepsilon(\mu_B,\delta,0) \p
\end{align}
It encodes the finite-temperature effects on the EoS, which are crucial to understand the role of temperature in proto-neutron stars and binary NS merger events. Generally, the thermal index shows a strong dependence on density and particle composition. 
A higher thermal index can lead to greater ejecta masses for a stiff EoS but can decrease for a softer EoS~\cite{Han:2025pho}. In our case the softening is induced by the population of the parity-partner degrees of freedom.

\begin{figure}
\centering
\includegraphics[width=1\linewidth]{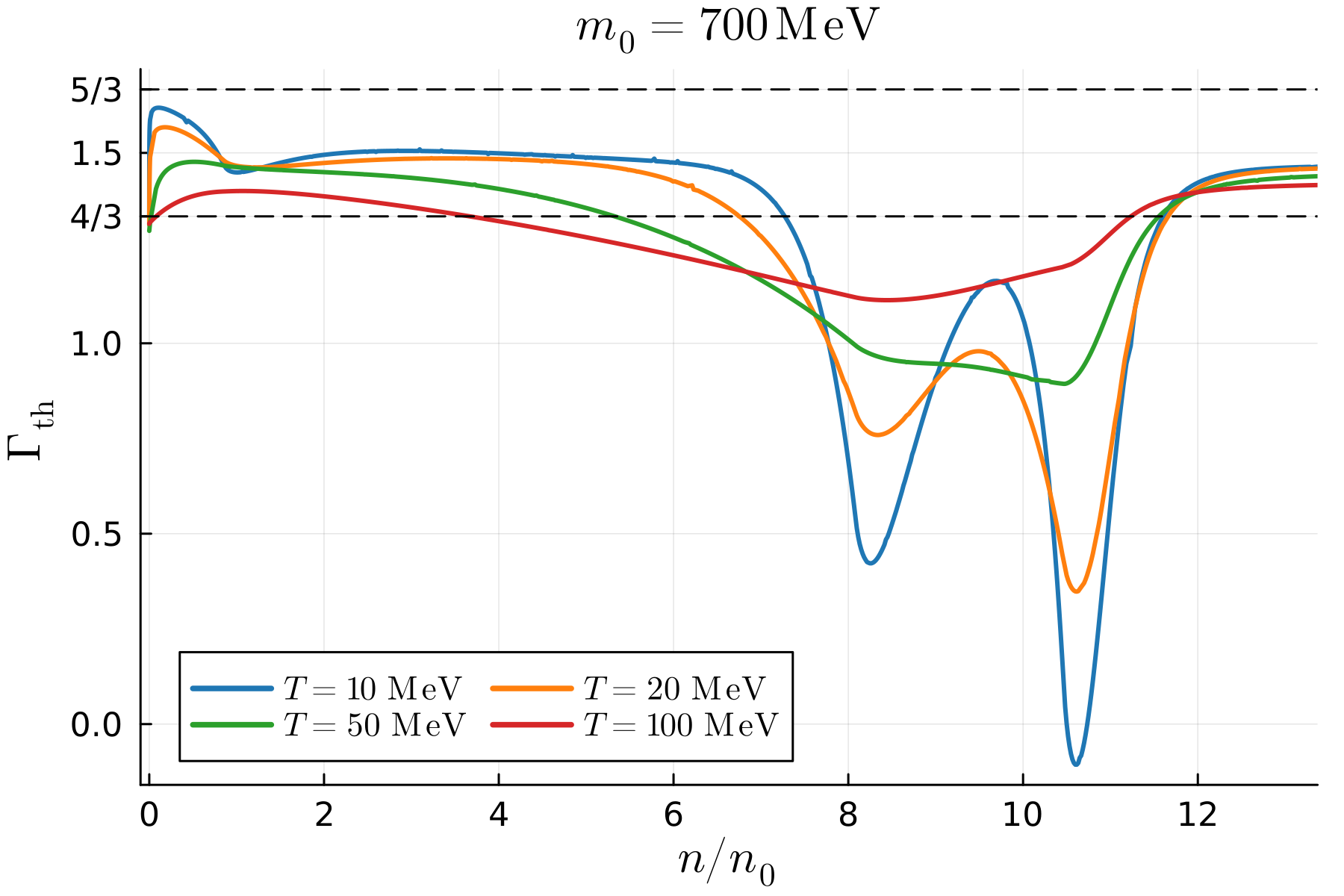}
    \caption{The thermal index $\Gamma_\mathrm{th}$ as a function of density in $\beta$-equilibrated NS matter for various temperatures, here exemplarily with $m_0=700$~MeV. 
    The horizontal lines indicate the free Fermi gas limit for $\Gamma_\mathrm{th}$: 5/3 for the non-relativistic gas and 4/3 for the ultra-relativistic one.
    }
    \label{fig:gamma}
\end{figure}

The results shown in Fig.~\Ref{fig:gamma}, are similar to what is obtained for the onset of hyperons in an $SU(3)$ EoS where a strong density dependence is observed \cite{Kochankovski:2025lqc}. The same is found in our case, albeit at much higher densities. The onset of the parity partners $n^*$ and $p^*$ (see Fig.~\ref{fig:abundances}) above $n/n_0\sim 8$ induces rapid variations in $\Gamma_\mathrm{th}$ which are most pronounced at low temperatures.

We note that the effect of temperature on the thermal index is to broaden the dips, until the double-dip structure is no longer recognizable.
For temperatures of 50~MeV or higher, this broadening causes the tail of the chiral transition to extend to lower densities, down to about six times the nuclear saturation density. These values of temperature and density are relevant in the context of neutron star mergers, raising the possibility that signatures of chiral restoration could be imprinted in such phenomena, e.g., in gravitational-wave signals.

In the low-density limit, the presence of electrons, which are ultra-relativistic at every temperature reported here, bring $\Gamma_\mathrm{th}$ close to the ultra-relativistic limit of 4/3. 

\section{Conclusions and Outlook}\label{sec:Conc}

In this work we have introduced a multiplicatively renormalizable MF formulation of the two‐flavor PDM in which the grand-canonical potential $\Omega$ is explicitly  expressed in an RG-invariant form. This RG-MF calculation is based on a parametrization of the mesonic effective potential $V(\phi^2)$, polynomial in $\phi^2 = \sigma^2 +\vec\pi^2$, for sigma meson and pions in terms of suitable powers of the parity‐doublet baryon masses $m_\pm(\phi)$.
In this formulation, the renormalization scale and the relevant renormalized dimensionless coupling are replaced by the RG-invariant scale parameter $\Lambda $ of the PDM, in a scheme where the reference scale in the renormalization condition is 
set to the chirally invariant PDM baryon mass $m_0$. 

For applications of the model to dense baryonic matter we have fitted (for a given $m_0$) the strengths of isoscalar $g_\omega$ and isovector $g_\rho$  vector repulsion between the baryons, and two dimensionless higher-order couplings $c_6$ and $c_8$ in our parametrization of the polynomial $V(\phi^2)$ (for the degree $n=6$ and $8$ interactions in units of $m_0$) to the properties of nuclear matter around saturation: the saturation density $n_0$, the binding energy per nucleon $E_B$, the nuclear compressibility $K_\infty$, the symmetry energy $E_{\rm sym}$ and the in‐medium chiral condensate $\sigma(n_0)$. The values for the slope parameter $L$ and the critical temperature $T_c$ of the nuclear LG transition resulting from the model calculation were not fitted but used as important sanity checks. 
Using this input from nuclear physics we have calculated chiral phase diagrams for symmetric and asymmetric nuclear matter for a wide range of temperatures as well as baryo- and isospin chemical potentials.

A major finding is that the inclusion of the fermionic VC to $\Omega$ moves the chiral transition to larger baryo-chemical potentials $\mu_B$ for symmetric nuclear matter or $\mu_n$ for pure neutron matter while substantially weakening any rapid low-temperature variations, especially for smaller values of $m_0$. At the same time, the position of the LG critical point as well as the zero-temperature phase diagram in the $(\mu_B,\mu_I)$-plane are essentially unaffected by the choice of $m_0$.  

For $\beta$-equilibrated cold neutron‐star matter, we find that chiral restoration, and the relative appearance of the negative‐parity partners of the neutron and proton, $n^*$ and $p^*$, lead to cusps and troughs in $c^2_s$. 
These occur at densities beyond $7\, n_0$ which, in all cases studied, are above the central density $n_\mathrm{max}$ of the star and thus irrelevant for the M-R relation. The values for the maximum mass tend to undershoot the observational $M_\mathrm{max}\geq 2M_\odot$ constraints, especially for larger values of $m_0$. From this we conclude that a more detailed description of the nuclear hard-core repulsion is needed. {In addition we showed that, at finite temperature, chiral restoration produces sizable variations in the thermal index $\Gamma_\mathrm{th}$ which persist at densities relevant for neutron-star mergers. It may therefore be possible that the partial restoration of chiral symmetry plays a role in such events}.


In view of our findings it is imperative to revisit the question of  what the role of strangeness is and whether hyperons appear in the inner core of NS. We wish to address this issue in future work by extending the RG‐consistent framework, discussed here, by promoting the PDM to the $SU(3)$-symmetric case, thus incorporating kaons and hyperons as well as the $U(1)_A$ anomaly.

To assess modifications of the finite-temperature
MF EoS it will be important to include fluctuations of the meson fields, especially the pion field, which can give sizable contributions to the thermal pressure, especially at high temperature \cite{Fore:2019wib}.
Including such mesonic fluctuations within the FRG is known to require going beyond the leading-order derivative expansion, i.e.~the local potential approximation employed previously \cite{Weyrich:2015hha}, to achieve a phenomenologically  successful description of the saturation properties of nuclear matter. Once this is available, realistic vector and axial-vector spectral functions to describe its electro-weak response could then be calculated from analytically continued FRG flows along the lines of Refs.~\cite{Jung:2019nnr,Tripolt:2021jtp}, in order to obtain quantitative predictions of thermal dilepton rates and neutrino emissivities for heavy-ion collision experiments and NS observations \cite{Wambach:2023tba}.
Meanwhile we plan to employ linear-response theory to evaluate small-amplitude vector and axial-vector fluctuations from Kubo relations in an expansion around the RG-invariant MF using an effective two-loop setup to account for the rescattering of pions in the medium \cite{Migdal:1978az}, which will be especially important for the low-frequency and low-momentum behavior of the electro-weak spectral functions in dense nuclear and neutron  matter. Work in this direction is in progress.

\begin{acknowledgments}
We thank Anton Andronic for discussions on the chemical freeze out at high-baryon densities and providing us with pertinent freeze-out data, Michael Buballa and Bengt Friman
for discussions on the generic features of the zero-temperature phase diagram, Hosein Gholami on RG consistency and for help with his open-source software used to solve the TOV equations, Ugo Mire for discussions and exchanges, especially on the numerical implementation of this work,
and J\"urgen Schaffner-Bielich for discussions on astrophysical constraints and effects of including strangeness. 
This work was supported by the Deutsche Forschungsgemeinschaft, project no.~315477589, the Collaborative Research Center CRC-TR 211, ``Strong-interaction matter under
extreme conditions.'' 
\end{acknowledgments}

\bibliography{kin.bib}

\end{document}